\documentclass[%
reprint,
 amsmath,amssymb,
 aps,
]{revtex4-2}

\usepackage{xcolor}
\usepackage{graphicx}

\begin{document}

\preprint{APS/123-QED}

\title{Simulating microswimmers under confinement with \\ Dissipative Particle (hydro)Dynamics}

\author{C. Miguel Barriuso G. and Jos\'e Mart\'in-Roca}
\affiliation{Departamento de Estructura de la Materia, F\'isica T\'ermica y Electr\'onica, Universidad Complutense de Madrid, 28040 Madrid, Spain}
 
\author{Valentino Bianco}%
\affiliation{Dpto. de Qu\'imica F\'isica, Facultad de Qu\'imica - Universidad Complutense de Madrid, 28040 Madrid, Spain}%

\author{Ignacio Pagonabarraga}
\affiliation{Departament de F\'isica de la Matèria  Condesada, Facultat de F\'isica - Universitat de Barcelona, Carrer de Mart\'i i Franquès, 1, 11, 08028 Barcelona, Spain}%
\affiliation{Universitat de Barcelona Institute of Complex Systems (UBICS)}
\affiliation{CECAM, Centre Européen de Calcul Atomique et Moléculaire, École Polytechnique Fédérale de Lasuanne (EPFL), Batochime, Avenue Forel 2, 1015 Lausanne, Switzerland.}

\author{Chantal Valeriani}
\affiliation{Departamento de Estructura de la Materia, F\'isica T\'ermica y Electr\'onica, Universidad Complutense de Madrid, 28040 Madrid, Spain}%
\affiliation{GISC - Grupo Interdisciplinar de Sistemas Complejos 28040 Madrid, Spain}


\begin{abstract}
In this  work we study  \textcolor{black}{microwimmers}, whether colloids or polymers, embedded   in bulk or in confinement. We explicitly consider hydrodynamic interactions and 
simulate the swimmers via an implementation inspired by the squirmer model. Concerning the surrounding fluid, we  develop  a Dissipative Particle Dynamics  scheme. Differently from the Lattice-Boltzmann technique, on the one side this approach allows us to 
properly deal not only with  hydrodynamics but also with thermal fluctuations.
On the other side, this approach  enables us to  study \textcolor{black}{microwimmers} with complex shapes, ranging from spherical colloids to polymers. 
To start with, we study a simple spherical colloid. 
We analyze the  features of the velocity fields of the surrounding solvent, when the colloid is a pusher, a puller or a neutral swimmer either in bulk or confined in a cylindrical channel.
Next, we characterise its dynamical behaviour by computing 
the mean square displacement and the long time diffusion when the active colloid is in bulk or in a channel (varying its  radius) and analyze the orientation autocorrelation function in the latter case. 
 While the three studied squirmer types are characterised by the same bulk diffusion, the  cylindrical confinement considerably modulates the diffusion  and the orientation autocorrelation function. 
 Finally, we focus our attention on a more complex shape: an active polymer. 
We first characterise the structural features computing its radius of gyration when in bulk or in cylindrical confinement, and compare to known results obtained without hydrodynamics.
Next,    we characterise the dynamical behaviour of the active polymer by computing its mean square displacement and the long time diffusion.
On the one hand,   both diffusion and radius of gyration  decrease due to the hydrodynamic interaction when the system is in bulk. 
On the other hand, the effect of  confinement  is to decrease the radius of gyration, disturbing the motion of the polymer and thus reducing its diffusion.
\end{abstract}

\maketitle

\section{Introduction}

Active Matter is a branch of Physics that focuses on the study of intrinsically out-of-equilibrium systems due to energy being constantly supplied, \textcolor{black}{converted into directed motion and} dissipated by  individual constituents. Active Matter is a field that has raised a lot of interest in the last decade, since it captures complex collective behaviours, often exclusively associated to living matter, and might enable a wide range of technological applications \cite{roadmap}. 
One of the paradigmatic systems of Active Matter consists of a suspension of active particles.  Active particles can be living (such as bacteria) or synthetic (such as active colloids). 
Active colloids are micron-size particles which self-propel through a medium by converting energy extracted from their environment into directed motion \cite{ActColAranson,ActColZottl}, with potential medical and technological applications \cite{de2020self,guzman2022, ActColEbbens,hortelao2018,Li2016, QIAN2020113548,Palacci2013}. The collective behaviour of systems constituted by a large number of these particles is rich and complex as shown by a series of recent numerical \cite{bianco,isele2015self,eisenstecken2016conformational,winkler2020physics,michieletto2020non,locatelli2021activity,das2021coil} and experimental \cite{deblais2020rheology} works, and in many cases cannot be ascribed solely to the particles motion since hydrodynamics due to the surrounding solvent might need to be taken into account \cite{martin2019active}. This is the case for  \textit{microswimmers} \cite{ElgetiWinklerGomper}, whose motion is an essential aspect of life. 

\noindent
Microswimmers  are usually ciliated \textcolor{black}{and/or flagellated} microorganisms that achieve propulsion thanks to the movement of their cilia located on their outer surface: for this reason one can consider them as self-propelled microorganisms.
\noindent
In the last few years  microswimmers have been intensively studied, being of interest in several interdisciplinary sciences.  Examples of living microswimmers are {\it Escherichia coli} bacterium, {\it Paramecium} or sperm cells, or algae (such as {\it Chlamidomonas} or \textit{Volvox}). Whereas examples of synthetic microswimmers are  Janus colloidal particles.
\noindent
When considering the effect of hydrodynamic interactions, numerical studies of a two dimensional suspension of self-propelled repulsive swimmers have demonstrated  that hydrodynamics affects, not only the phase behaviour of a dense suspension \cite{29}, as suggested  by Ishikawa \cite{30} in an early work, but also the dynamics of transient clusters at lower densities \cite{31}. \textcolor{black}{Moreover, other theoretical, numerical and experimental results have also revealed the importance of hydrodynamics in these systems \cite{schwarzendahl_2018,schwarzendahl_2019,stenhammar_2017,liu_2021}}.

\noindent
To model microswimmers,  Blake and Lighthill proposed the  so called \textit{squirmer}  model \cite{lighthill,blake}. 
The squirmer model reproduces  the induced hydrodynamic flow around a spherical swimmer while preserving the main features of the active stresses generated by it \cite{39}.
The spherical squirmer particle mimics  the effect of the cilia on the fluid as a prescribed slip velocity tangential to the surface. The described mechanism is the one that leads to the   swimmer's propulsion.  A squirmer is  characterized by two modes accounting for its swimming velocity and its active stress. 
Depending on the active stress, it is possible to classify squirmers as  pushers (e.g., {\it E. coli}, sperm), pullers (e.g., {\it Chlamydomonas}) and neutral  (e.g., {\it Paramecium}) swimmers \cite{Lauga_2009,theers}. The squirmer model has been expanded for complex swimmers, such as non-spherical swimmers \cite{Theers7372} and explicitly  ciliated microorganisms \cite{Lauga_2009}.

\noindent
Besides mimicking  the swimmer's behaviour, it is important to choose a model to mimic the features of the surrounding fluid.  
The applicability of atomistic algorithms (Molecular Dynamics-like) to simulate the fluid is limited, since they only allow to study  short time and length scales (few hundreds of nanoseconds and  few tens of nanometers).  To explore  longer length/time scales,  more relevant for living swimmers,   atomistic methods become computationally inefficient. Thus, one might consider  mesoscopic methods, that bridge the gap  between the microscopic   and the macroscopic continuum scale \cite{Groot}. These methods  span  longer length and time scales:  from several nanometers to micrometers and from nanoseconds to microseconds. The most renown mesocopic numerical models used to simulate fluids that fully consider hydrodynamic interactions are Lattice-Boltzmann \cite{sauro2001},  Multiparticle Collision Dynamics \cite{MCD1,MCD2,MCD3} and Dissipative Particle Dynamics \cite{DPD}. The Lattice Boltzmann (LB) approach consists in describing the solvent in terms of the density of particles with a given velocity at a node of a given lattice. The discretized velocities join the nodes and prescribe the lattice connectivity \cite{46}. The LB model reproduces the dynamics of a Newtonian liquid of a given shear viscosity $\eta$. Relevant hydrodynamic variables are recovered as moments of the one-particle velocity distribution functions. The total force and torque the fluid exerts on a particle embedded in it are obtained by imposing that the total momentum exchange between the  particle and the fluid nodes vanishes. Since a Lattice Boltzmann code is computationally expensive, from a practical point of view it is possible to parallelize it using  Message Passage Interface to exploit the excellent scalability of LB on supercomputing facilities \cite{49}.
In the Multiparticle Collision Dynamics (MPCD) approach \cite{MCD1,MCD2,MCD3} 
a fluid is represented by N point particles with continuous positions  and velocities. The particle dynamics proceeds in two steps: streaming and collision. 
During the streaming step,  particles move ballistically. Whereas in the collision step particles  interact  locally via an instantaneous stochastic process, that could be based on   stochastic rotation dynamics  with angular momentum conservation \cite{theers}. 
For this purpose, the simulation box is partitioned into cubic collision cells.
Within MPCD Galilean invariance is ensured, together with thermal fluctuations. 
The algorithm conserves mass, linear, and angular momentum on the collision cell level, which gives rise to hydrodynamics on large length and long time scales.
Dissipative particle dynamics (DPD) is one of the most efficient mesoscale coarse-grained \textcolor{black}{approaches} for modeling soft matter systems. DPD was originally proposed by Hoogerbrugge and Koelmann \cite{DPD} as an off-lattice, momentum conserving, Galilean invariant mesoscopic method, the coarse-grained dynamics of which obeys the Navier-Stokes equations and preserve hydrodynamics. Later on, Espanol and Warren \cite{espanolwarren} reformulated the DPD model in terms of stochastic differential equations. DPD consists in modified Langevin equations that operate between pairs of particles interacting via  three different forces: conservative, dissipative and random (thermal) forces. The DPD model has already been  used to model complex colloidal suspensions, such as  
 proteins \cite{Wei234902} or red globules in blood \cite{Fedosov2011}.
 
\noindent
\textcolor{black}{Along with hydrodynamics, confinement also plays a major impact on the dynamics of microswimmers. Their interaction with bounding walls is different depending on the type of microswimmer we are dealing with and can lead to different transport and aggregation phenomena. As of today, this facts have been studied theoretically \cite{blake_wall,spagnolie_lauga_2012,lauga_ch4}, numerically \cite{zhu_2013,holmLBsquirmer,starkMPC} and experimentally \cite{Lauga_2009,berke_2008}. While in these works the study focuses mainly in the interaction of microswimmers with plane boundaries, other types of confinement also display relevant features, for instance the effect of porous media in bacterial suspensions has also been reported \cite{Kurzthaler2021}. Finally, of course it is worth noting that the effect of confinement is not only limited to active matter systems but also plays a role in a wide range of systems, such as passive hard spheres \cite{Mandal2014}.}

\noindent
In the present work we propose to model suspensions of \textcolor{black}{microwimmers} with DPD hydrodynamics inspired by the squirmer model. When the agent is a sphere \textcolor{black}{we choose a \textit{raspberry}-like structure \cite{rasp_lobaskin,rasp_graaf_2015,rasp_graaf_2016}} and we will directly consider the squirmer model. Whereas when the agent is a polymer, we will build the polymer as a chain of monomers, and treat each monomer similarly, but not rigorously, as a squirmer.
To properly deal with hydrodynamics, we will mimic the surrounding fluid via DPD interactions, using an in-house extension of the LAMMPS \cite{LAMMPS} open source package \textcolor{black}{implementing appropriate reaction forces on the swimmer's particles that balance the forces exerted on the fluid and enable its propulsion \cite{nash_rat-hydro_2010,menzel_ddft_2016}}. Our choice is motivated by the fact that 
differently from LB \cite{sauro2001}, DPD easily allows to take into account thermal fluctuations and to simulate colloids with complex shapes (not only spherical). \textcolor{black}{Moreover, it is also easier to control compressibility and Schmidt number in DPD than MCPD. Hence, in DPD it is easier to control the appropriate dynamic regime that couples the solvent and solute dynamics. Although this is  not the focus of his paper, DPD also allows for a more  thorough control of the phase diagram of the solvent and how to deal with fluid phase coexistence.} Firstly we study the dynamical behaviour of either \textcolor{black}{microwimmer}  in bulk.  In the case of  active colloids, we establish the flow fields surrounding the particle and compute their diffusion, comparing pushers, pullers and neutral swimmers. 
In the case of   active polymers, besides the dynamics we also study its conformational features. Next, we  confine either \textcolor{black}{microwimmer}  in a cylindrical channel, and unravel the effect of hydrodynamics as compare to the equivalent systems where hydrodynamics is not present. 
For each system we  explore different  Reynolds and P\'eclet numbers.
The Reynolds number \cite{Chisholm233} is the ratio of inertial  to viscous forces within a fluid subjected to relative internal motion: this number measures the amount of turbulence of the solvent  in the system. The P\'eclet number \cite{StarkPeclet,bianco} is defined as the ratio of the rate of advection of a physical quantity by the flow to the rate of diffusion of the same quantity. This number quantifies the degree of activity of \textcolor{black}{microwimmers}. 

\noindent
The manuscript is organised as follows. In section \ref{sec:methods} we describe the relevant physical quantities and the technical details of the implementation. We first describe the DPD method  to simulate the solvent (sec. \ref{subsec:solv-DPD}),  implemented within the LAMMPS open source numerical package \cite{LAMMPS}. Next, we present the two \textcolor{black}{microwimmers} under study: the active colloid (sec. \ref{subsec:colloid}) and the active polymer (sec. \ref{subsec:polymer}). In sec. \ref{subsec:colloid}, we introduce the {\it raspberry}-like active colloid  (fig. 1.a) in bulk and when interacting with a cylindrical surface (fig. 1.c). In section \ref{subsec:polymer}, we report the active polymer (fig. 1.b), as in ref. \cite{bianco}, in bulk and under cylindrical confinement (fig. 1.d). 
The way we implemented hydrodynamics is reported in Section 2.4, being  the same for both active objects embedded in a DPD solvent. In the same section we  characterize the physical quantities of a fluid such as the kinematic viscosity ($\nu$) and the solvent diffusion coefficient ($D_{sol}$), and parameters to quantify the activity of the colloid/polymer embedded in a fluid, such as the Reynolds number and the P\'eclet number. 
Finally, in Section 2.5 we report the analysis tools used to study the \textcolor{black}{microwimmers} in bulk or under confinement. In section \ref{sec:results} we present the results obtained, first for the colloid (sec. \ref{subsec:colloid}) and then for the polymer (sec. \ref{subsec:polymer}). In section \ref{sec:discussion} we discuss the results and comment on future avenues. 

\section{Materials and Methods}\label{sec:methods}

In this work we  study an active colloid and an active polymer embedded in a fluid solvent either in bulk or confined inside a cylindrical channel. We simulate the active colloid as a spherically-shaped collection of particles merged together by rigid interactions. Whereas the active polymer is built as a chain of monomers glued together by harmonic interactions that enable their relative movement. The rest of the interactions  are the DPD-like interactions between any two particles, the hydrodynamic force-field that enables the agents' propulsion and, in the case of the confined polymer, a repulsive (WCA-like) potential between  channel (particles) and  polymer/solvent particles.

\subsection{Modeling the solvent with Dissipative Particle Dynamics}\label{subsec:solv-DPD}
\noindent
Our system consists of \textcolor{black}{microwimmers} embedded in a solvent, where hydrodynamics is explicitly taken into account. 
The fluid surrounding the \textcolor{black}{microwimmer} is simulated as a collection of individual particles interacting via Dissipative Particle Dynamics \cite{DPD}. According to DPD, below a given cutoff $r_c,$ the force acting on the $i$-th solvent particle consists of three contributions, 
\begin{equation}
    \vec{F}_i = \sum_j  \left( \vec{F}_{ij}^C + \vec{F}_{ij}^D + \vec{F}_{ij}^R \right)  \hat{r}_{ij} \quad \quad \text{if} \quad r<r_c,
\end{equation}
being $\hat{r}_{ij}=(\vec{r}_i-\vec{r}_j)/r_{ij}$  the inter-particle unitary direction between the $i$-th and $j$-th particles. The conservative term is $\vec{F}_{ij}^C = A \, w(r_{ij})$, where $A$ is the amplitude and $w(r_{ij})=1 -r_{ij}/r_c$ a weighting factor varying between 0 and 1 as in ref. \cite{Groot}. The dissipative contribution reads $\vec{F}_{ij}^D = - \gamma \, w^2(r_{ij}) \, \left(\hat{r}_{ij} \cdot \vec{v}_{ij} \right)$ with friction coefficient $\gamma$. Finally, the thermal contribution  $\vec{F}_{ij}^R = \sigma \, \alpha \, w(r_{ij}) \, /\sqrt{\Delta t}$ is a random force, where  $\alpha$, is a Gaussian random number with zero mean and unit variance, $\Delta t$ the chosen time-step for the time integration and $\sigma = \sqrt{2 \,k_B T\, \gamma }$ is related to the mean of the random force via fluctuation-dissipation, being $T$ the temperature of the system.

\noindent
In the current work, we implement the DPD solvent  via the LAMMPS open source package \cite{LAMMPS}, setting the time step to  $\Delta t/\tau =10^{-2}$ for the simulations of the active colloid and $\Delta t/\tau =5\cdot 10^{-3}$ for the simulations of the active polymer. In both systems, we choose an equilibration time of $\sim 10^4$ steps, while the production run is of the order of $10^6$ steps. The number of solvent particles for the system contaning the active colloid in bulk is  $N=10125$, distributed in a cubic simulation box of $L=15$. In cylindrical confinement, depending on the channel radius $R_\text{cyl}=\{3.5,4.5,5.5,6.5\}$ the number of solvent particles is $N_{sol}=\{3464,5726,8553,11946\}$, respectively, and the channel length is fixed to $L_\text{cyl}=30$. The number of solvent particles for the polymer system in bulk is around $N_{sol}=24000$ distributed in a cubic simulation box of $L=20$. In the polymer confined case, with channel radius $R_\text{cyl}=6$ and length $L=50$, the number of solvent particles is around $N_{sol}=17000$. For all  simulations, the mass of the solvent particles is fixed to $m=1$ and the numerical density to $\rho_\text{sol} =3$. The characteristic length scale for all our simulations is the DPD cutoff distance between solvent particles $r_c\equiv r_c^{ss}=1$, the mass scale is fixed by the mass of one solvent particle $m=1$, and for the time scale we fix $\tau=1$. Following ref. \cite{Groot}, the DPD interaction parameters between solvent-solvent particles are set to $A^{\text{ss}}=25.0$, $\gamma^{\text{ss}}=4.5$ and $r_c^{\text{ss}}=1.0$ (see tables  \ref{tab:DPDparam-colloid} and \ref{tab:DPDparam-polymer} in the following subsections). The physical properties of a DPD fluid depend on its viscosity \cite{Groot} that can be computed from the Green-Kubo relation \cite{kubo1957} for the stress autocorrelation function (zero-shear viscosity). Later on we will discuss our choice for the fluid's viscosity. Whereas the DPD parameters used for each \textcolor{black}{microwimmer} are  reported in their corresponding sections. 

\noindent
\textcolor{black}{It is worth noting that when confining the DPD fluid in a cylinder we observed concentric ring-shaped density fluctuations in the vicinity of the wall at $T=0.1$. These fluctuation have been previously observed and can lead to undesirable effects \cite{pivkin_2006}. For this reason we chose $T=1$ for all our confined simulations in which these fluctuations were not observed.}

\subsection{Colloids in bulk and in confinement}

\begin{figure}[h!]
\centering
    \includegraphics[width=\columnwidth]{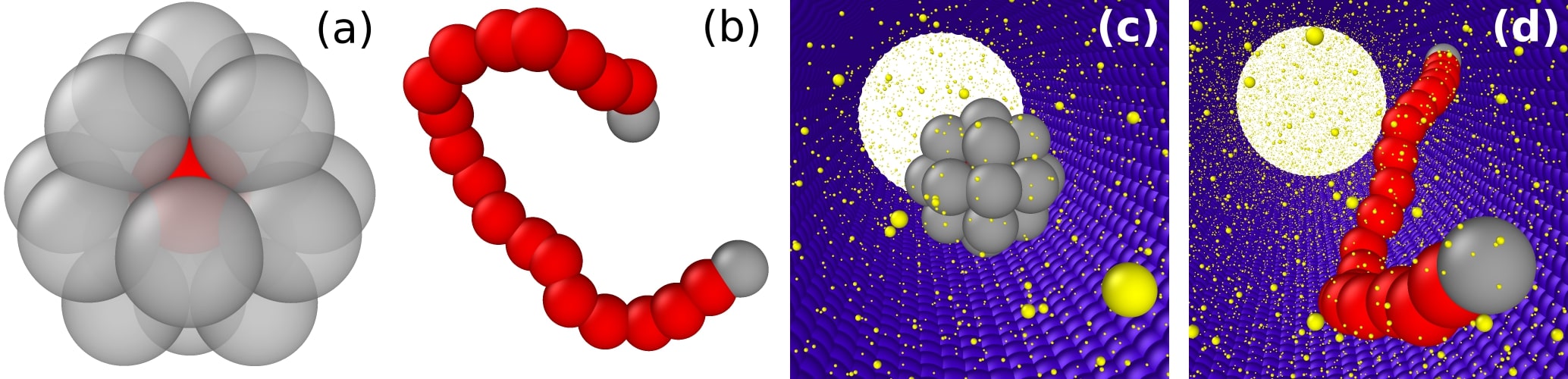}\hspace{1cm}
    \caption{\textbf{a:} A {\it raspberry}-like active colloid composed of 18 \textit{filler} particles and one \textit{thruster} particles at the center. Note that in these figures the filler particle radius is scaled down to $1$ for better visibility, but in all simulations we use $r_c^\text{sf}=2$ as the solvent-filler DPD cutoff in order to obtain a more spherical colloid and to prevent the solvent particles from stepping into the colloid.  \textbf{b:} An active polymer composed of 20 beads. \textbf{c:} An active colloids in a cylindrical confinement. \textbf{d:}  An active polymer in a cylindrical confinement. Red: \textit{thruster} particles. Gray: \textit{filler} particles. Blue: wall particles composing the confining channel. Yellow: solvent particles. }
    \label{fig:col-pol-snaps}
\end{figure}

\noindent
To study a spherical  squirmer, we build a {\it raspberry}-like \cite{rasp_lobaskin,rasp_graaf_2015,rasp_graaf_2016} colloid made of 19 particles rigidly bonded. In fig. \ref{fig:col-pol-snaps}.a, we represent the active colloids, consisting of  one particle (the \textit{thruster} particle)  located at the center of the sphere and the remaining 18 (\textit{filler} particles)  evenly distributed on the surface of a sphere of radius $R_\text{col}$ around the center particle. \textcolor{black}{The reason for choosing this structure has been guided by simplicity, balancing the number of particles and sphericity, and is inspired by previously  proposed models for complex colloids \cite{rasp_lobaskin,rasp_graaf_2015,rasp_graaf_2016}. The reason to consider only one thruster particle is because such an approach is enough to generate activity with a minimal disturbance on the geometrical properties of the swimmer. The shell of passive particles is necessary to control the dimension and shape of the colloid, rendering it an extended body that has an orientation.} The propulsion mechanism of the thruster particle will be explained in the next section, when detailing the implementation of the hydrodynamic interactions.

\noindent 
The orientation of the colloid is defined by the ``active axis'' identified by three chosen co-linear particles. This axis is also the symmetry axis of the force field we will apply to the solvent, and thus will define the colloid's direction of propulsion. All  particles belonging to each colloid interact via DPD:  1) with the solvent, 2) with particles belonging to other colloids and 3) with  particles building  the channel. However,  particles belonging to each colloid do not interact between them (so their overlap does not cause any trouble), except for the rigid interactions that keep them glued together. In table \ref{tab:DPDparam-colloid} we report the chosen DPD parameters for all interactions between particles: solvent-solvent, solvent-colloid, solvent-cylinder, colloid-colloid, colloid-cylinder. 
\begin{table}[h!]
\centering
\begin{tabular}{ |c|c|c|c|c|c| }
\hline
& $SOL$-$SOL$ & $SOL$-$COL$ & $SOL$-$CYL$ & $COL$ -$COL$  & $COL$ -$CYL$ \\
\hline
$A$         & 25.0  & 25.0  & 100.0 & 25.0  & 25.0  \\ 
$\gamma$    & 4.5   & 4.5   & 4.5   & 4.5   & 4.5   \\ 
$r_c$       & 1.0   & 2.0   & 1.0   & 2.0   & 2.0   \\ 
\hline
\end{tabular}
\caption{DPD parameters used to study a spherical colloidal squirmer for ($COL$) embedded in a solvent ($SOL$), in bulk  or in a cylindrical confinement ($CYL$). All parameters are in reduced DPD units.}
\label{tab:DPDparam-colloid}
\end{table}

\noindent
In Section 2.4 we will describe different squirmer models, such as pushers, pullers and neutral swimmers, each one characterised by a different velocity field in the surrounding fluid. In order to check whether the {\it raspberry}-like colloid reproduces the features of the different squirmers, we compute the velocity fields and compared them to those reported for the different squirmers in ref. \cite{starkMPC}.

\noindent
Having studied the active colloid in bulk, we study its physical behaviour when confined in cylindrical environments of different radii. The cylinder is composed of DPD  overlapping particles, properly aligned along the $x$ axis at given  angles. Overlapped DPD particles  are left out of the time integration and their DPD interactions are switched off thus they can be used to model a wall.
Particles are first evenly distributed along a circumference in the $yz$-plane and then this circumference is repeated through the $x$-axis. The separation of the particles is chosen so that the roughness of the inner surface of the cylinder is the same along the angular and longitudinal directions. 
Periodic boundary condition (PBC) are applied along the longitudinal direction ($x$ axis). Particles'  interaction parameters are reported in table \ref{tab:DPDparam-colloid}. Choosing DPD interactions for modelling the collisions with the channel allows us to maintain a large time step $\Delta t = 10^{-2}$. Due to the softness of the DPD interactions, we have appropriately set the DPD parameters for the channel particles  to avoid  leaking of solvent particles through the channel wall. Moreover, DPD enables adding a friction between the solvent and the channel wall. In our case, we have tested that for high enough values of $\gamma$ we are able to simulate Poiseuille flow. However, for our study we have decided to explore low values  of $\gamma$, which correspond to the implementation of slip boundary conditions at the channel's surface.

\subsection{Polymers in bulk and in confinement}

\noindent
Following ref. \cite{bianco}, we model the active polymer as a chain of active monomers. As shown in fig. \ref{fig:col-pol-snaps}.b, each of the $N_b$ monomers is composed by a single {\it thruster} DPD particle, except the head and tail monomer. \textcolor{black}{Since the first (last) particle of the polymer does not have previous (posterior) neighbors, no force is applied on them. Alternatively, one could consider that the activity direction is extrapolated from the neighbor monomer in the chain. However, this alternative approach  will not affect the main results and features described in the manuscript. Previously proposed models for active polymers  have also taken this approach \cite{bianco}.}
Monomers  are held together to their first neighbours  via a harmonic potential $V_{\rm {harmonic}}(r) = K (r-r_0)^2$, acting between  {\it thruster} particles of the connected beads separated by a distance $r$, with $K=30k_BT/r_c^2$, being $r_0\equiv 1.5 \, r_c$. Since all interaction between particles are soft (DPD-like), we can choose $dt=10^{-2}$ as the time step to integrate the equations of motion. As in ref. \cite{bianco}, we assume that all  monomers are active  apart from the first and the last (in grey in fig. \ref{fig:col-pol-snaps}). An active force $\textbf{F}_{a,i}$  acts on each \textit{thruster} monomer at ${\bf r}_{i}$. The force is characterised by a constant magnitude $F_a$ and a direction of  ${\bf r}_{i+1} - {\bf r}_{i-1}$ parallel to the polymer backbone tangent, being ${\bf r}_{i+1}$ and ${\bf r}_{i-1}$ the position vectors of the \textit{thruster} particles first neighboring monomers.

\noindent
To characterise the bulk properties of an active polymer, we study a dilute system of 4 active polymers in a box with edge $L=20$ at a solvent density of $\rho=3$. Care must be taken if the volume fraction of polymers is not low enough, since polymers  might interact between each other via hydrodynamics. In our case we avoid this by working with a polymer volume fraction that is always lower than 5\%. 

\noindent 
To study the effects of confinement we embed the active polymer and the solvent in a cylindrical channel with periodic boundary conditions along the axial axis. The cylinder consists of $N_c=24415$  frozen  WCA-like particles that interact with the DPD particles (solvent and polymers) via a WCA-like potential 
\begin{equation}
 V_{\rm LJ}(r) = 
\begin{cases}
4 \epsilon 
\left[ \left(\frac{\sigma}{r} \right)^{12} -
  \left(\frac{\sigma}{r} \right)^{6} \right] 
+ \epsilon; &{\text {for}}\,\, r<2^{1/6}\,\sigma, \cr
0; &{\text {for}}\,\, r \geq 2^{1/6}\,\sigma,
\end{cases}
 \label{eq:LJ}
\end{equation}
where $\epsilon$ is the unit of energy and $\sigma$ represent the channel's particle diameter set to $\sigma=r_c=1$. In all simulations we set $k_B T=1.0$ (Lennard-Jones units). 
Cylinder particles  are located  close enough  to avoid  DPD solvent particles to cross the cylinder's wall.

\noindent
The chosen values for the DPD parameters are reported in table \ref{tab:DPDparam-polymer} for all interactions between particles: solvent-solvent, solvent-polymer, solvent-cylinder, polymer-polymer, polymer-cylinder.   
\begin{table}[h!]
\centering
\begin{tabular}{ |c|c|c|c|c|c| } 
\hline
& $SOL$-$SOL$ & $SOL$-$POL$ & $SOL$-$CYL$ & $POL$-$POL$ & $POL$-$CYL$ \\
\hline
$A$         & 25.0  & 25.0  & 25.0  & 25.0  & 0.0  \\ 
$\gamma$    & 4.5   & 4.5   & 4.5   & 4.5   & 0.0   \\ 
$r_c$       & 1.0   & 1.0   & 1.0   & 1.0   & 0.0   \\ 
$\epsilon$  & 0.0   & 0.0   & 0.0   & 0.0   & 1.0  \\ 
\hline
\end{tabular}
\caption{DPD parameters used to study an active polymer ($POL$) embedded in a solvent ($SOL$), in bulk  or in a cylindrical confinement ($CYL$). All parameters are in reduced DPD units.}
\label{tab:DPDparam-polymer}
\end{table}

\noindent
We should stress the fact that when dealing with active colloids or active polymers we have chosen to simulate the cylindrical channel in a different way.  In the former case, the channel has been simulated by means of particles interacting via DPD, as explained earlier. Whereas in the latter case,  the channel has been built using particles interacting via a repulsive WCA potential, to compare with ref. \cite{bianco}.

\subsection{Swimming induced by hydrodynamics}

\noindent
In order to numerically consider full hydrodynamic interactions between the \textcolor{black}{microwimmers} and the surrounding solvent,  we prescribe a force field for the solvent particles surrounding the thruster particles (the red particles in fig. \ref{fig:col-pol-snaps}). 

\noindent
When dealing with a spherical squirmer, the usual approach consists in prescribing tangential velocities to the solvent particles at the swimmers surface \cite{lighthill}. 
Note that  we have not followed the usual squirmer approach.
In our case, tangential solvent \textit{forces}, instead of velocities, are prescribed over a hydrodynamic active volume $\Gamma_H$ around the colloid, instead of just at the colloid's surface. This approach is more general since it enables the possibility of studying different agent shapes and inertial effects, which are present in many active systems \cite{lowen_inertial}. 

\noindent
In this study we only consider  axisymmetric force fields (eq. \ref{ec:axi-hydro-force-field}). We choose the hydrodynamic region $\Gamma_H$ as a spherical shell around the thruster particles of inner and outer radii $R_c$ and $R_H$, respectively. The expressions of the force fields considered for the colloid and the polymer are reported in what follows (see the appendix for more details). The general expression for an axisymmetric force field that vanishes everywhere except inside the aforementioned spherical shell is given by,
\begin{equation}\label{ec:axi-hydro-force-field}
    \boldsymbol{f}(r,\theta) = \big[
    f_r(r,\theta) \boldsymbol{\hat{e}}_r + 
    f_\theta(r,\theta) \boldsymbol{\hat{e}}_\theta\big] P_{R_c,R_H}(r) \
\end{equation}
where $r$ is the distance from the thruster  to the solvent particle, $\theta$ the angle between the colloid's orientation vector $\hat{\boldsymbol{e}}$ and the solvent position vector, $\hat{\boldsymbol{e}}_r$ and $\hat{\boldsymbol{e}}_\theta$  are the radial and tangential unitary vectors with respect to the colloid frame of reference and $P_{R_c,R_H}(r)=\Theta(r - R_c)\Theta(R_H - r)$ is a pulse function in the radial dimension which defines the spherical shell.

\begin{figure}[h!]
    \centering
    \includegraphics[width=\columnwidth]{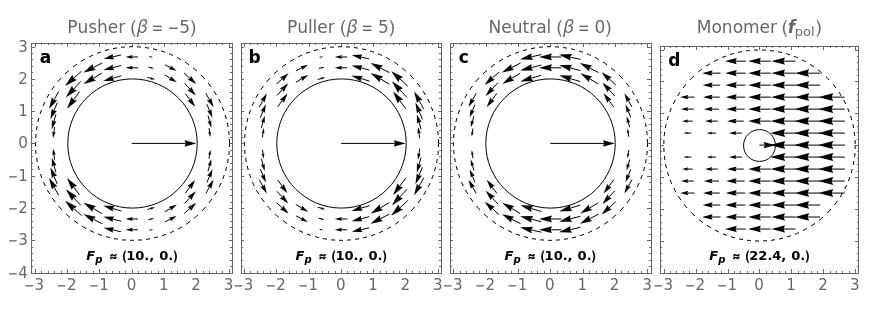}
    \caption{Examples of 2D hydrodynamic redistribution force fields for all the studied cases: a pusher (panel a) with $\beta=-5$, a puller (panel b)  with $\beta=5$, a neutral squirmer (panel c)  with $\beta=0$ and a monomer of the active polymer (panel d). These  correspond to a section of the 3D field passing through the equator of the redistribution sphere. The arrow inside the central colloid indicates the direction of the propulsion force $F_p$, with magnitude indicated in each panel. \textcolor{black}{Here, the hydrodynamic region $\Gamma_H$ would be the area contained between the solid and dashed circles.}}
    \label{fig:redistribution}
\end{figure}

\noindent
In order to have more control over the propelling force, we normalize the force field over the hydrodynamic region $\Gamma_H$,
and multiply by a factor $F_p$. $F_p$ is the input parameter for the magnitude of the self-propelling force. Thus the hydrodynamic force field that will be applied to the solvent reads,
\begin{equation}
    \boldsymbol{f}_H(r,\theta)=F_p\frac{\boldsymbol{f}(r,\theta)}{N},\quad \text{where} \quad N=\left|\int_{\Gamma_H}\boldsymbol{f}(r,\theta)\,\text{d}V\right|.
    \label{ec:force_on_solvent}
\end{equation}
Since in our case we are dealing with a discrete fluid (made of solvent particles), the $i$-th solvent particle will feel a force,
\begin{equation}
    \boldsymbol{f}_H^{\,i}=F_p\frac{\boldsymbol{f}(r_i,\theta_i)}{N},\quad \text{where} \quad     N=\bigg|\sum_{j\in\Gamma_H}\boldsymbol{f} (r_j,\theta_j) \, \Delta V\bigg|
    \label{ec:force_on_solvent_discrete}
\end{equation}
where the sum is taken over all the solvent particles that are inside $\Gamma_H$ and $\Delta V = r_c^3 = 1$. Similarly to the squirmer model, here we only consider the two first surface modes of the polar component, $f_\theta(r,\theta)$, while the radial component is neglected $f_r(r,\theta)=0$  (see appendix sec. \ref{subsec:hydro_details} for the details).
Thus, the force field becomes,
\begin{equation}
    \boldsymbol{f}(r,\theta) = (B_1\sin\theta+B_2\sin\theta\cos\theta)P_{R_c,R_H}(r)\boldsymbol{\hat{e}}_\theta\equiv \boldsymbol{f}_\text{col},
\end{equation}
for which $N=(R_H^3-R_c^3)B_1\pi^2/4$. In this way, the total propulsion force (which is precisely the integral appearing in eq. \ref{ec:force_on_solvent}) experimented by the colloid is just $F_p$. $B_2$ controls the force dipole contribution to the force field and thus wether we are dealing with pushers ($B_2<0$), neutrals ($B_2=0$) or pullers ($B_2>0$). Because of this formulation, $B_1$ plays no role and will be fixed to $1.0$ from here on.
As in the squirmer model, 
we define $\beta =B_2/B_1$ as the active stress parameter that  controls the type of squirmer  (see fig. \ref{fig:redistribution}). Under the assumption of Stokes flow (low Reynolds number), it is reasonable to think that the velocity field of the solvent particles will resemble that of the squirmer model\footnote{Under the same Stokes flow assumption, the colloid's propulsion velocity $v_p$ can be computed from the self-propulsion force $F_p$ via the Stokes law $F_p=6\pi\eta R_c v_p$. However, this assumption may not hold in some cases that are also worth studying. For these cases, we ``measure'' $v_p$ as the time averaged projection of the colloids velocity $v_\text{col}$ over its orientation axis $v_p = \langle\boldsymbol{v}_\text{col}\cdot \hat{\boldsymbol{e}}\rangle_t$ as will be explained later on.} \cite{starkMPC,lighthill}.

\noindent
Now we need to deal with the reaction force that is exerted on the colloid which will result in its thrust. Moreover, since an active colloid is an extended rigid object we would like to preserve the torque that may arise due to density fluctuations or interactions with other objects.  The reaction thrust force, $\boldsymbol{f}_T$, is applied on the nearest colloid particle (thruster or not) to each of the solvent particles and it is equal and opposite to the redistributed force on that solvent particle: 
\begin{equation}
    \boldsymbol{f}_T^{\,k}(r,\theta)=-\boldsymbol{f}_H^{\,i(k)}(r,\theta)
\end{equation}
where $i(k)$ represents the nearest solvent particle to the $k$-th colloid particle.

\noindent
At each step, we implement the following algorithm: \begin{enumerate}
    \item \textcolor{black}{Starting from a reference microswimmer,} we identify the neighboring solvent particles around the swimmer's thruster particles  located  between the swimmer's radius ($R_\text{col}$ for the colloid; $r_c$ for the polymer) and the ``hydrodynamic'' radius $R_H$.  
    \item We compute the force field $\boldsymbol{f}_H$  in Eq. (\ref{ec:force_on_solvent})  at each of the neighbors positions, consistently with the swimmer's orientation. The norm of the total distributed force is also computed.
    \item For each neighbor:
    \item[] \begin{enumerate}
                \item[3.1] We apply the corresponding normalized force.
                \item[3.2] We find the nearest agent particle and apply the same and opposite force.
            \end{enumerate}
\end{enumerate}

\noindent
In this way self-propulsion is achieved, while linear and angular momenta are locally conserved at each step. This procedure enables physically realistic modeling of the propulsion mechanism of a wide range of self-propelled systems, both living and artificial.

\noindent
In  case of the active polymer, we have considered a constant field modulated by $\cos\left(\theta/2\right)$ for each thruster monomer,
\begin{equation}
    \boldsymbol{f}(r,\theta) = -\cos\left(\frac{\theta}{2}\right)P_{R_c,R_H}(r)\,\boldsymbol{\hat{e}}\equiv \boldsymbol{f}_\text{pol},
\end{equation}
where $\boldsymbol{\hat{e}}=\cos\theta\,\boldsymbol{\hat{e}}_r-\sin\theta\,\boldsymbol{\hat{e}}_\theta$ is the self-propulsion direction of the thruster particle. In this case, a reaction force that provides thrust to the agent is applied on each thruster particle. This force is equal and opposite to the total force distributed among the solvent particles in each step. Since in this case we are dealing with a flexible object that has many thruster particles, we need not to worry about finding the nearest agent particle, since this is already taken care of, as the force that each thruster particle redistributes is equal and opposite to the one that is exerted on it.

\noindent
In figure \ref{fig:redistribution} we can see the hydrodynamic force fields, $\boldsymbol{f}$, we have used in this work. The continuous and dashed circumferences represent the inner ($R_c$), and outer ($R_H$) radius respectively and define the region $\Gamma_H$ where redistribution occurs. As an example, in this figure the propulsion force is computed as the surface integral of the vector field inside this region, $\boldsymbol{F}_p = -\int_{\Gamma_H} \boldsymbol{f}_H(r,\theta)\,d\boldsymbol{S}$, in this case computed in 2D. Since the force field is asymmetric there exists a net propulsion force that provides thrust to the agent. In the case of the polymer (d), each polymer bead (or monomer) acts as a small colloid with its own redistribution field, so in this case $R_c=r_c$ would represent the beads radius, i.e. the thickness of the polymer.

\noindent
To conclude, the total force experienced by an agent particle consists of the following contributions
\begin{equation}
    \boldsymbol{F}=\boldsymbol{F}_C+\boldsymbol{F}_D+\boldsymbol{F}_R+\boldsymbol{F}_T
\end{equation}
where $\boldsymbol{F}_T$ is the total thrust force, computed as the sum of all the reaction forces on each colloid's particle $\boldsymbol{F}_T=\sum_k \boldsymbol{f}_T^k$. 
\noindent
It is worth noting that while in this study we have restricted ourselves to axisymmetric force fields, the code implementation is made for general force fields, allowing for example azimuthal flows, like those of the Volvox algae \cite{Drescher2009}.

\subsubsection{Quantifying activity}

\noindent
To characterise a \textcolor{black}{microwimmer} in the solvent, we will define dimensionless numbers such as the Reynolds number and the P\'eclet number. For this, we will need to establish the viscosity $\eta$ of the fluid. $\eta$ can be numerically computed in a DPD fluid, as recently shown in ref. \cite{Panoukidou2021}, or estimated via a mean field, as in  Warren and Groot \cite{Groot}. In our work, we follow the second approach, according to which the DPD solvent kinematic viscosity $\nu$, defined as $\frac{\eta}{\rho}$,  can be computed as 
\begin{equation}
\label{ec:solv-kinem-viscosity}
    \nu=\frac{\eta}{\rho}=\frac{D_{sol}}{2} + \frac{2\pi\gamma\rho  r_c^5}{1575}
\end{equation}
where the diffusion coefficient $D_{sol}$ is
\begin{equation}
 D_{sol}=\frac{45k_BT}{2\pi\gamma\rho r_c^3}.
 \label{eq:diffusion_mean_field_DPD}
\end{equation}
 For more details, see Warren and Groot \cite{Groot}. 
Note that in MPCD the viscosity can be computed as $\eta = 16.05 \, \sqrt{m_0k_B T}/a_0^2$, where $m_0$ and $a_0$ are the mass and the size of the cell used in MPCD algorithm. See \cite{StarkPeclet,Noguchi2005,Kikuchi2003,Tuzel2003} for more details.

\noindent
Once we know the viscosity, we compute the Reynolds number and the P\'eclet number. 
The Reynolds number quantifies the amount of inertial versus viscous forces acting on an object that moves in a fluid, 
\begin{equation}\label{ec:Reynolds}
    \text{Re}=\frac{v_p r_c}{\nu}
\end{equation}
where $r_c=1$ is the solvent characteristic length and $v_p$ is the \textcolor{black}{microwimmer}'s propulsion velocity. For both the active colloids and the active polymer, the velocity is the one of the center of mass. 

\noindent
\textcolor{black}{The P\'eclet number is an adimensional number used to quantify the measure of the activity. It is directly proportional to the self-propulsion speed and to the reorientation time \cite{JoseJCP}. The P\'eclet number has been used in bulk suspensions of Active Brownian Particles \cite{Stenhammar_2014, fang_2022} to quantify particles' activity. Differently from ABP, when dealing with swimmers (as in our case) the rotational dynamics of the colloid arises from interaction with the fluid and it is not prescribed (such as in ABP). Therefore we introduce  a  P\'eclet number based on the propulsion velocity, $v_p$. In this way one should expect that, for a given set of parameters, the reorientation time increases with the propulsion velocity and thus with the P\'eclet. In other words, } the P\'eclet number is defined to describe the degree of activity in the system as the ratio between the self-propulsion of the \textcolor{black}{microwimmer} and a \textcolor{black}{diffusion} scale $\mathrm{Pe}=v_p\,\tau_r/\sigma$. However, different works have shown different definitions for this number. 
We define the P\'eclet number for the colloid following ref. \cite{StarkPeclet} as,
\begin{equation}\label{ec:Peclet_col}
   \text{Pe}_{col}=\frac{v_p R_\text{col}}{D_{col}}
\end{equation}
here $D_{col}=k_B T/6\pi\eta R_{col}$ is the estimated diffusion coefficient of the colloid and $R_\text{col}$ is the colloid radius that is fixed to 2 for all our simulations (with the exception of the flow fields shown in figure \ref{fig:flow-fields_bulk} in which $R_\text{col}=3$ was chosen for better visibility). We have studied the following ranges $F_p\in[0,50]$, $v_p\in[0,1.25]$, $\mathrm{Pe}\in[0,156]$ and $\mathrm{Re}\in[0,12]$. As mentioned earlier, for some parameters we cannot assume that we are in Stokes flow conditions, so we should not use the relation  $v_p=F_p\,/\,6\pi\eta R_c$ for computing the colloids propulsion velocity, this is why we ``measure'' it as $v_p = \langle\boldsymbol{v}_\text{col}\cdot \hat{\boldsymbol{e}}\rangle_t$. The values obtained are shown in fig. \ref{fig:prop_vel} of the appendix sec. \ref{subsec:prop_vel}. Note that it was found that the propulsion velocity of the colloid, and thus the $\mathrm{Pe}$ and $\mathrm{Re}$, are not always linear with the propulsion force $F_p$. Moreover, they change whether we are dealing with pusher, neutral or puller squirmers. In fig. \ref{fig:prop_vel} we show the different propulsion velocities found (and their corresponding $\mathrm{Pe}$ and $\mathrm{Re}$) for each type of squirmer. However, in all simulations presented, we remain in the range $F_p\in[0,\,50]$ where the separation between the $v_p$ values for different squirmer types is not so dramatic and the behaviour does not depart too much from linearity.

\noindent
For the polymer we follow the ref. \cite{bianco} and define the P\'eclet number as
\begin{equation}
    \text{Pe}_{pol}=\frac{F_p \, r_{c}}{k_B T}
\end{equation}
where $r_{c}=1$ represents the characteristic length of the monomers. The polymer's lengths, $N_b$, studied lay in a range between 40 and 100. For this particular cases we have explored values $\mathrm{Pe}=\{0.01, 0.1, 1.0\}$ that correspond with Reynolds numbers in the laminar regimen, around Re$\approx 0.3$.

\subsection{Analysis tools}

\noindent
In order to characterise our systems, we compute both structural and dynamical features. 
Concerning the active colloid, we first  establish  the velocity field of the solvent surrounding the swimmer to characterise the nature of each spherical squirmer (whether pusher, puller or neutral). Next, we  study its dynamics by computing the mean square displacement, and estimate the effective diffusion coefficient from its long time behaviour. When dealing with the colloid in confinement we also analyze the orientation autocorrelation function (OACF) that supplies information about the reorientation time of the colloid.
Concerning the active polymer, we first characterise how activity affects its structural features by computing the radius of gyration. Next, we  study its dynamics by computing the mean square displacement of the center of mass and again estimate the effective diffusion coefficient from its long time behaviour. 

\noindent
\textbf{Velocity fields.} For computing the solvent velocity fields around the colloid we run simulations of a fixed colloid in the center of the box pointing to the positive $x$-axis. Then, we perform a binning of the simulation box and average the velocities of the solvent particles inside each bin, finally we also take ensemble and time averages in the stationary state. The velocity fields shown in fig. \ref{fig:flow-fields_bulk} correspond to a slab that has the same height as the colloid ($D_\text{col}$). The arrows represent the $xy$-projection of the full 3D velocities.

\noindent
\textbf{MSD.} Concerning dynamical features, we compute the mean square displacement
\begin{equation}
    \mathrm{MSD}(t)= \langle \, \left[\mathbf{r}_{\mathrm{cm}}(t)-\mathbf{r}_{\mathrm{cm}}(0)\right]^{2} \,\rangle
\end{equation}
Where $\mathbf{r}_{\mathrm{cm}}$ indicates the position of the center of mass of the colloid/polymer. The  average is taken over several colloids/polymers. The long time behaviour of the MSD, corresponds to the diffusion coefficient $D$, $\mathrm{MSD}(t)= 6 D t$. It is worth noting that when  confinement takes place in a cylinder with a small radius,  it might be  better to consider the system as one dimensional,  thus   $\mathrm{MSD}(t)= 2 D t$. However, this is not  our case since we consider that the agents have sufficient space to diffuse in the transverse directions. This leads to a more straightforward comparison between the different systems.

\noindent
\textbf{OACF} The orientation autocorrelation function is computed for the colloid in confinement to asses the impact of the confinement in the rotational diffusion (or equivalently, the reorientation time) of the colloid.
\begin{equation}
    \mathrm{OACF}(\Delta t) = \frac{1}{N_{\Delta t}}\sum_{i=0}^{N_{\Delta t}} \hat{\boldsymbol{e}}(t_i)\cdot \hat{\boldsymbol{e}}(t_i+\Delta t)
\end{equation}
Here $\Delta t =n\,dt$ where $dt$ is our base time step. The scalar product of the orientation at a given time $\hat{\boldsymbol{e}}(t_i)$ with itself at a delayed time $\hat{\boldsymbol{e}}(t_i+\Delta t)$ is averaged over the intervals of length $\Delta t$, starting at all the possible $t_i$'s, that fit into the total simulation time $T_\mathrm{sim}=N_\mathrm{tot}dt$. So there would be $N_{\Delta t}=N_\mathrm{tot}-n+1$ intervals of the same length in the full simulation interval for a given $n$. 

\noindent
\textbf{RoG.} The radius of gyration $R_g$ for the active polymer is computed according to the relation,
\begin{equation}
    R_{g}^{2} = {\frac {1}{N}}\sum _{k=1}^{N}\left(\boldsymbol{r}_{k}-\boldsymbol{r}_{\mathrm{cm}}\right)^{2}, 
\end{equation}
where $\boldsymbol{r}_{\mathrm{cm}}$ is the position of the center of mass of the polymer,  $\boldsymbol{r}_{k}$ is position of the $k$ {\it thruster} particle and  $N$ is the number of bead of the polymer


\section{Results}\label{sec:results}
In what follows we present the results obtained for both \textcolor{black}{microwimmers}, either in bulk or in cylindrical confinement. We  start with the simplest object: the spherical squirmer (Section 3.1) characterising  its hydrodynamic features (Section 3.1.1) and its dynamical properties (Section 3.1.2). When confined in a cylindrical channel, we also compute its orientation autocorrelation function (Section 3.1.3). 
Next, we  study the more complex-shape active polymer (Section 3.2), characterizing its structural (Section 3.2.1) and dynamical (Section 3.2.3) properties, compare our results with the passive and Brownian counterpart.
\subsection{Active colloids}\label{subsec:colloid}

\subsubsection{Flow Fields}
\noindent
To start with, we present our results for a spherical squirmer and study the velocity fields for the pusher, the puller, and the neutral swimmer.
\begin{figure}[h!]
\begin{center}
\centering
    \includegraphics[width=\columnwidth]{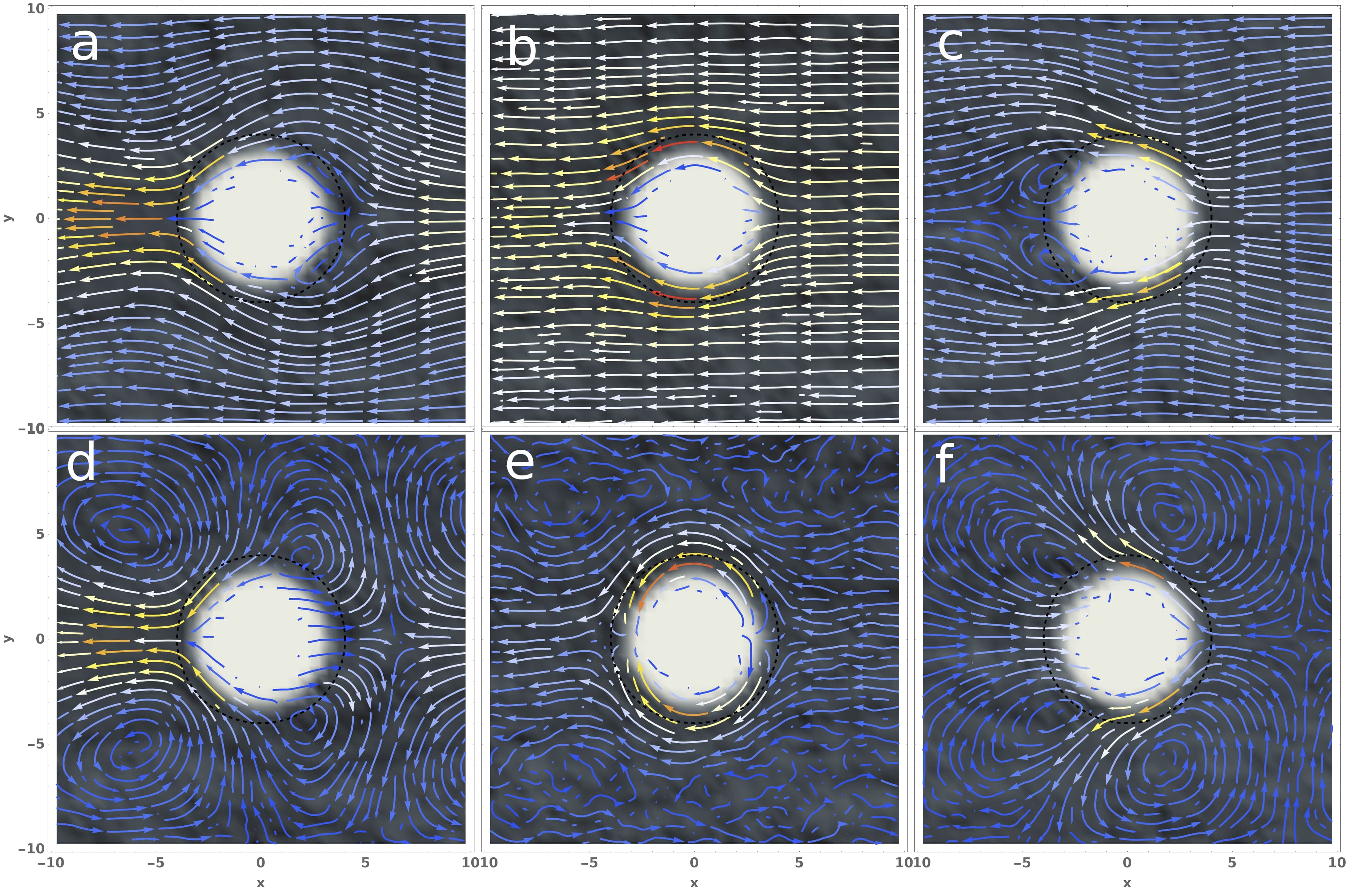}
    \caption{Sections of the velocity fields in the lab frame (bottom row) and moving with the colloid (top row), for the pusher (a, d), neutral (b, e) and puller (c, f) squirmer. 
    surrounded by $\sim 10^4$ fluid particles and swimming to the right. The colour of the background is the averaged density of fluid particles, the colour of the arrows shows the fields magnitude. Here the colloid radius is $R_c=3$ and $F_p=100$ in order to obtain a clearer flow field. These values produce $\mathrm{Pe}\approx\{818,\,655,\,573\}$ and $\mathrm{Re}\approx\{25,\,20,\,17\}$ for the pusher, neutral and puller squirmers respectively.}
    \label{fig:flow-fields_bulk}
\end{center}
\end{figure}

\noindent
 Fig. \ref{fig:flow-fields_bulk} displays  the velocity flow fields for this three squirmers computed as explained in the previous section:  pusher (a, d), neutral (b, e) and puller (c, f) squirmer.  Comparing our results with the typical flow fields expected for squirmers (e.g. ref. \cite{starkMPC}) the flow fields reported in Fig.  \ref{fig:flow-fields_bulk} are not so symmetrical, in the case of the puller and pusher lab frames (figs. \ref{fig:flow-fields_bulk}.d and \ref{fig:flow-fields_bulk}.f).
 The four characteristic vortices of the flow field when periodic boundary conditions are present \cite{holmLBsquirmer} seem to be shifted to the negative $x$-direction, compressing the two at the front and stretching the two at the back. In the same way, in the relative frame, we can see smaller swirls than usual at the front of the pusher (fig. \ref{fig:flow-fields_bulk}a) and somewhat elongated ones at the back of the puller (fig. \ref{fig:flow-fields_bulk}.c). In the case of the neutral swimmer, the characteristic source dipole of the lab frame (fig. \ref{fig:flow-fields_bulk}.e) is completely compressed against the swimmers surface, and some turbulent flow is appreciated at the edges of the $y$-dimension of the section. All these deviations from the usual flow fields are ascribed to inertial effects of the fluid stemming from the high Reynolds number present in our simulations\cite{Chisholm233}. In Appendix  \ref{subsec:low_reynolds} we show the flow fields for a different set of parameters (lower Reynolds number, at $\mathrm{Re}\approx 0.1$) for which we find a more typical squirmer flow field \cite{starkMPC,holmLBsquirmer}. \textcolor{black}{In ref. \cite{holmLBsquirmer} the authors also comment on an analytical solution and state that it is indistinguishable from the one found in their simulations. An analytical solution for the flow field without PBCs in terms of a source dipole, a force dipole and a source quadrupole is also provided in ref. \cite{lauga_ch4} following a different approach but compatible with the usual derivation by Blake followed by \cite{holmLBsquirmer}.  Ref. \cite{Chisholm233} studies in detail how the Reynolds number affects the flow fields around squirmers.}

\begin{figure}[h!]
    \centering
    \includegraphics[width=\columnwidth]{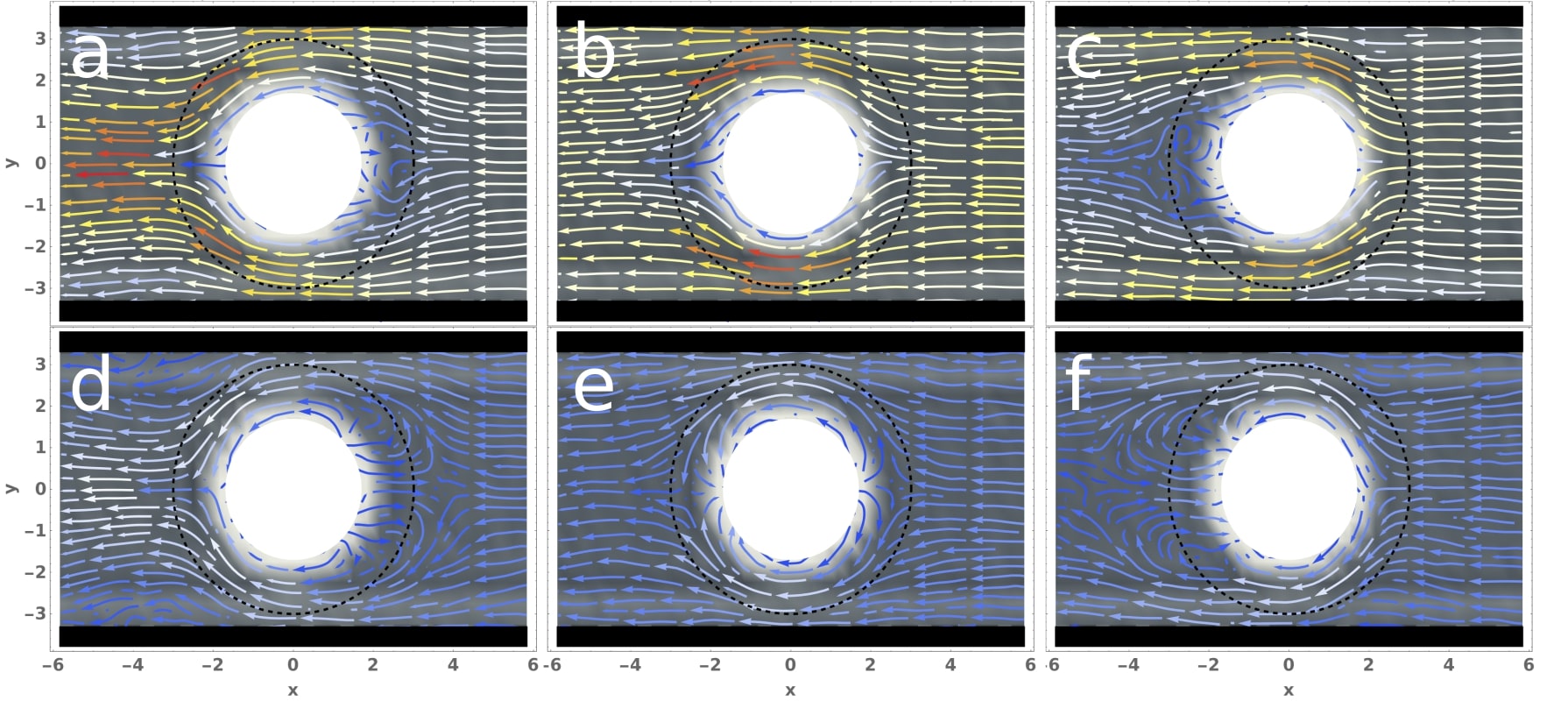}
    \caption{Solvent flow fields of the colloid confined in an cylindrical channel ($R_\text{cyl}=3.5$) in the lab frame (bottom row) and moving with the colloid (top row), for the pusher (a, d), neutral (b, e) and puller (c ,f) squirmer. Here the colloid radius is $R_c=2$ and $F_p=50$. These values produce $\mathrm{Pe}\approx\{90,\,76,\,61\}$ and $\mathrm{Re}\approx\{3.9,\,3.3,\,2.7\}$ for the pusher, neutral and puller squirmer respectively.}
    \label{fig:flow-fields_chan}
\end{figure}

\noindent
Confining the colloid inside a cylindrical channel has a drastic impact in the solvent flow fields (fig. \ref{fig:flow-fields_chan}), since the channel walls change the boundary conditions of the fluid. 
For the pusher and the puller in the absolute frame (figs. \ref{fig:flow-fields_chan}.a and \ref{fig:flow-fields_chan}.d) we observe  that the two vortices at the back and front respectively have disappeared, while the other two (at the front of the pusher and at the back of the puller) seem to have retracted to a closer position directly in front of the pusher and behind the puller. \textcolor{black}{A similar  damping of vorticity has already been reported in ref. \cite{ignacio_confined_colloid} and may be attributed to the suppression of the fluid's long-wavelength modes due to the confinement \cite{bocquet-barrat_confined-fluid}.}
In the relative frame of reference, the swirls have also contracted further, and it is now difficult to distinguish them from just turbulent flow. In the case of the neutral swimmer (figs. \ref{fig:flow-fields_chan}.b and \ref{fig:flow-fields_chan}.e) the flow fields do not differ that much with respect to the ones encountered in bulk, with the exception that now there are no turbulent regions at the edges of the flow field.

\subsubsection{Diffusion}

\begin{figure*}[t!]
\begin{center}
    \centering
    \includegraphics[width=0.75\textwidth]{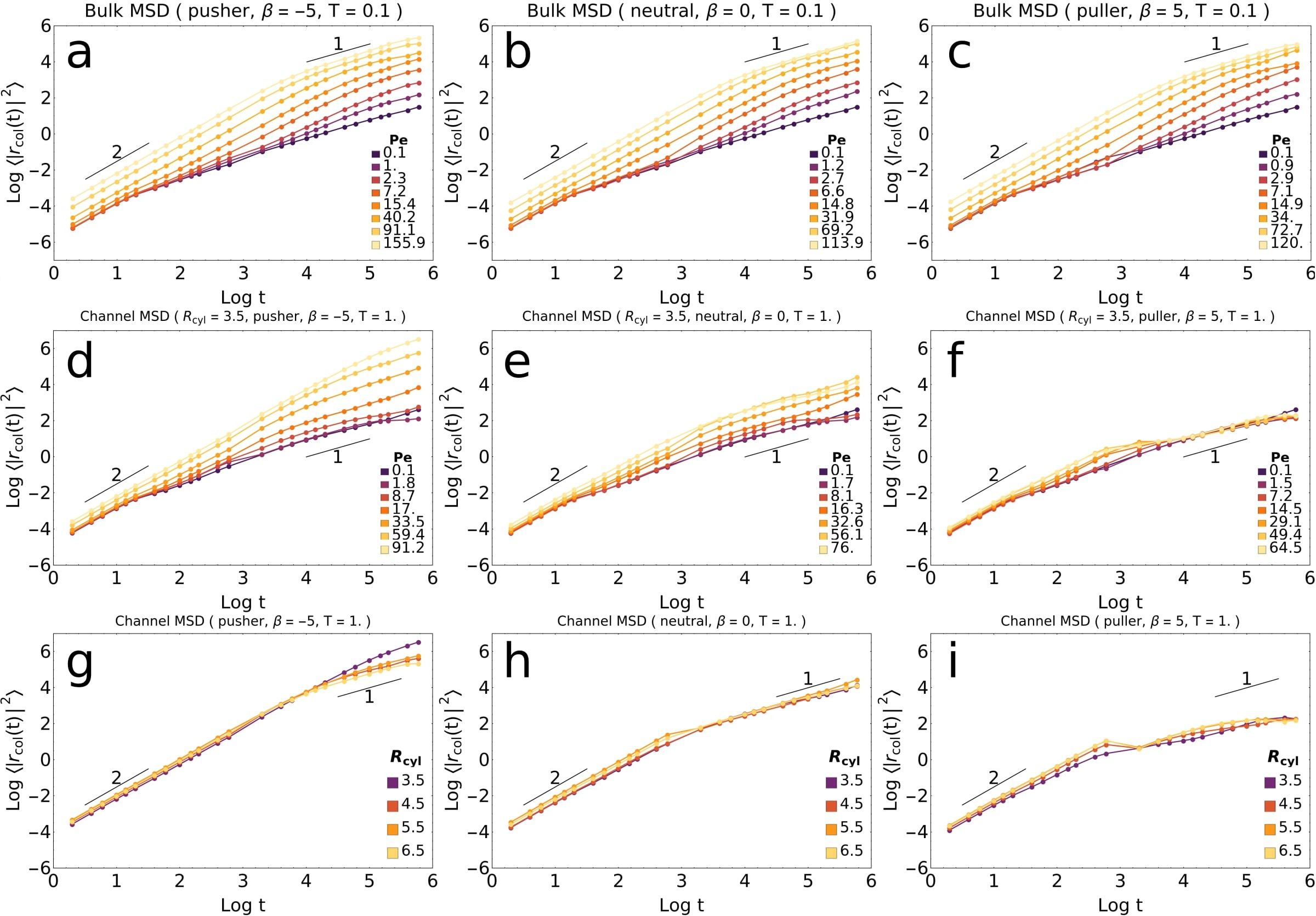}
    \caption{MSDs for the three squirmer types studied. \textbf{Left column:} pusher (panels a, d and g). \textbf{Center column:} neutral (panels b, e and h). \textbf{Right column:} puller (panels c, f and i). \textbf{Top row:} bulk (panels a, b and c). \textbf{Center row:} in cylindrical confinement for different $\mathrm{Pe}$'s for the smallest channel radius $R_\text{cyl}=3.5$ (panels d, e and f). \textbf{Bottom row:} for different channel radii for the highest P\'eclet numbers available corresponding to the highest thrust force $F_p=50$ (panels g, h and i). }
    \label{fig:msds}
\end{center}
\end{figure*}

\noindent
When dealing with a colloidal squirmer in bulk, we study its dynamical features by estimating the long time diffusion coefficient normalised by the diffusion of a passive colloid in bulk 
via  the center of mass mean square displacement, as explained in Section 2.5., for the three types of squirmers (figs. \ref{fig:msds}a-c).

\begin{figure}[h!]
\begin{center}
    \centering
     \includegraphics[width=\columnwidth]{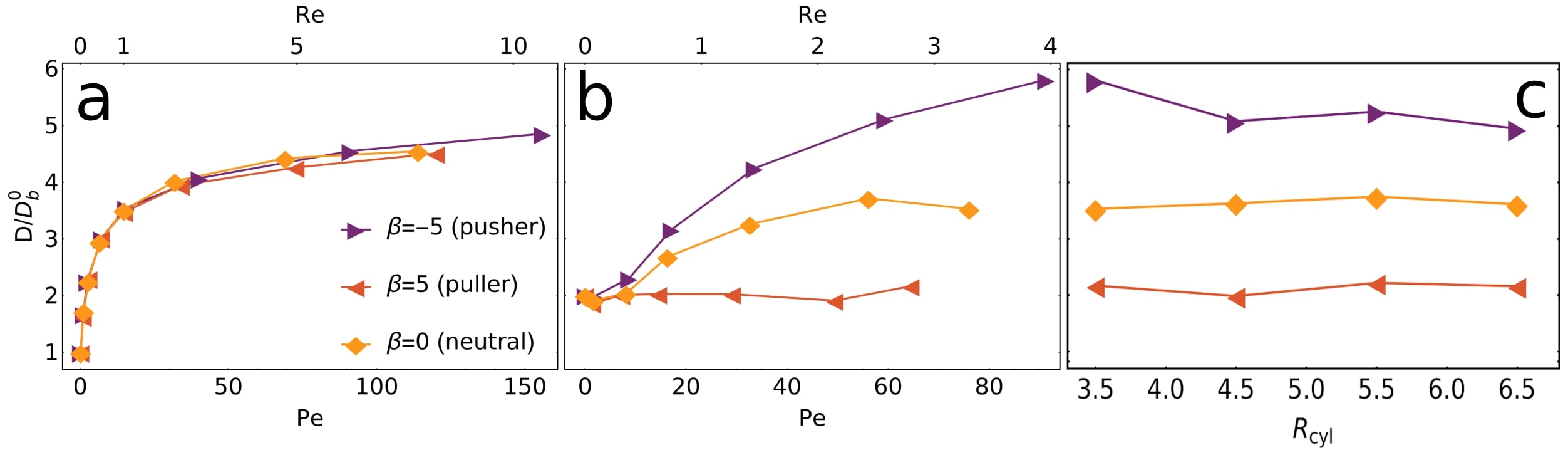}
    \caption{Measured diffusion as a function of the P\'eclet number for the active colloid in bulk (a), in confinement for the narrowest channel $R_\text{cyl}=3.5$ (b) and as a function of the cylinder radius for the highest P\'eclets (c) for the three squirmer types studied. We normalize by the diffusion of a passive colloid in bulk $D_b^0=0.032179$. In the two first plots a secondary horizontal axis shows the corresponding Reynols number.}
    \label{fig:diff_col}
\end{center}
\end{figure}

\noindent
The top row of panels in fig. \ref{fig:msds} represent the MSD for a bulk dilute suspension of pushers (a), neutrals (b) and pullers (c). 
Their long time behaviour corresponds to the diffusion coefficient reported in fig. \ref{fig:diff_col}.a 
 From the results presented, it is reasonable to conclude that the three types of squirmer diffuse almost the same for the ranges of P\'eclet numbers studied. As expected, the diffusion of the three of them increases  when  increasing their thrust force and thus their P\'eclet number. 
  \noindent
 In fig. \ref{fig:diff_col}.a it is worth noting that as we increase the thrust force and thus the P\'eclet and Reynolds numbers, the diffusion behavior changes significantly, when we are in the range of $\mathrm{Re}\ll 1$ the diffusion increases significantly while we increase the $\mathrm{Pe}$, when we approach $\mathrm{Re}\approx 1$ the increase in diffusion is dampened reaching what seems to be a saturation as $\mathrm{Re}\gg 1$.

\noindent
\noindent
The middle and bottom row of panels in fig. \ref{fig:msds} represent the MSD for a confined dilute suspension of pushers (a), neutrals (b) and pullers (c). 
The middle panels study the dynamics of swimmers in a channel with the smallest radius, while varying the Peclet number for pushers (d), neutrals (e) and pullers (f). 
The bottom panels study the dynamics of swimmers at the highest Peclet in a channel with varying  radius for pushers (g), neutrals (h) and pullers (i).
When we confine the active colloid inside a cylindrical channel the symmetry between pushers and pullers is lost. \textcolor{black}{Pushers will tend to reorient parallel to the wall, while pullers will do so perpendicularly. In this way, mobility of pushers should be increased while pullers should be more prone to getting ``stuck'' at the wall,} which is  reflected in the MSD curves in figs. \ref{fig:msds}d-f. \textcolor{black}{This is a well known behavior for the interaction of squirmers with walls \cite{starkMPC,spagnolie_lauga_2012}, that has been used to explain the accumulation of certain microorganisms at surfaces \cite{berke_2008}. The coupling between  a  microswimmer close to the wall and the solvent is affected because a portion of its hydrodynamics region (see fig. \ref{fig:redistribution}) lies outside the cylinder, where no solvent particles are present. Although this effect can contribute an additional torque, the volume of this excluded region is small compared with the rest of the hydrodynamic region, and it is not seen to affect the qualitative features of the hydrodynamic  coupling between the microswimmer and the confining wall. Moreover, when studying the diffusion for different channel radii (fig. \ref{fig:diff_col}.c), if this effect had a relevant contribution, we would expect to see a clear modulation of the diffusion with the varying radius for all types of squirmers, since the greater the radius, the smaller the excluded portion of the hydrodynamic region.}

\begin{figure*}[t!]
\begin{center}
    \centering
    \includegraphics[width=0.8\textwidth]{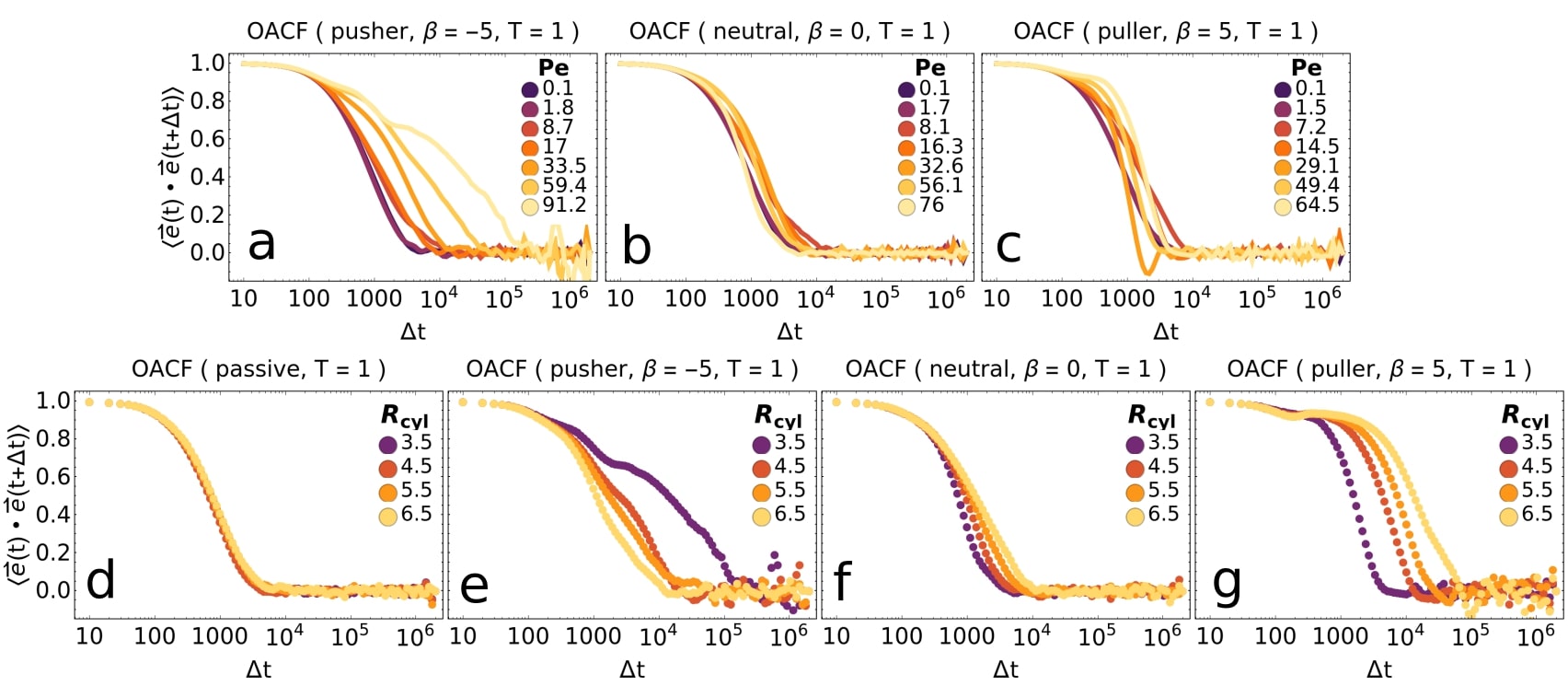}
    \caption{Auto-correlation function for the colloids orientation vector. \textbf{Top:} for the smallest cylinder radius $R_\text{cyl}=3.5$ and all the P\'eclet numbers studied for pusher (a), neutral (b) and puller (c). \textbf{Bottom:} for all the studied cylinder radii for a passive colloid (d)for which $F_p=0$, $\mathrm{Pe}=0.1\approx 0$ and the highest P\'eclet numbers available (all corresponding to $F_p=50$) for the three types of squirmers: pusher (e), puller (f) and neutral (g).}
    \label{fig:OACF_chan}
\end{center}
\end{figure*}

The same information will be recovered when plotting the  OACF for each system (as will be shown in figs. \ref{fig:OACF_chan}e-g) curves).

\noindent
As shown in the MSD curves (figs. \ref{fig:msds}d-f), for the pusher we detect a slight increase at large times, while for the puller the curves collapse showing a significant decrease in its motility at large times for all studied P\'eclet numbers .
This is due to the wall-facing effect described previously, which is consistent not only with the decrease of motility, but also with the apparent independence of the diffusion with the P\'eclet number.  
 The shape of the MSDs curves for the neutral swimmer (fig. \ref{fig:msds}e) also follow from this argument. The neutral squirmer gains its thrust force symmetricaly between its front and back.
 Therefore,  it propels on its front more than the pusher but less than the puller, and propels on its back more than the puller but less than the pusher. The fact that this system is  between the two is confirmed by the MSDs curves. The diffusion curves (fig. \ref{fig:diff_col} a,b,c) show more clearly what we have just addressed.

\noindent
In fig. \ref{fig:diff_col}c we report the normalized diffusion for highest P\'eclet number of the three types of squirmers in confinement as a function of the channel radius. The major effect of varying the channel radius occurs for the pusher, while the puller and neutral squirmer's diffusion seems to remain unaffected by it (in the studied range). 
This is coherent with the wall-facing argument previously described. The diffusion is a long time property, while for the studied radii the colloid reaches the channel wall at much shorter time scales. 
Therefore once the colloid has reached the wall, it might get stuck due to the wall-facing effect regardless of the channels radius.


\subsubsection{Orientation aturocorrelation function}

\noindent
Finally, we compute  the  orientation auto-correlation function (OACF) when active colloids are confined in a cylindrical channel, as depicted in fig. \ref{fig:OACF_chan}. The OACF   measures  the rotational diffusion (or equivalently, the reorientation time) of a colloid, i.e. for how long the colloid retains its swimming direction before it is randomized by fluctuations.

\noindent 
The top row of fig. \ref{fig:OACF_chan} represents the OACF for the system confined in the smallest cylinder, when varying the P\'eclet number. Whereas the bottom row represents the OACF for an active colloid propelling at the highest P\'eclet number and confined in cylinders with different radii.
 In the case of a pusher (fig. \ref{fig:OACF_chan}a) we  detect a clear increase of the reorientation time with increasing $\mathrm{Pe}$. 
 This is expected for any non-chiral active particle which increases its $\mathrm{Pe}$ by increasing its propulsion force \cite{JoseJCP}.
 Moreover, due to the wall-rebound argument discussed previously, this effect could be amplified. 
When dealing with the neutral  (fig. \ref{fig:OACF_chan}b) and the puller (fig. \ref{fig:OACF_chan}c) squirmers, the interpretation is less  clear. 
It  seems that in both cases starting from the lowest $\mathrm{Pe}$ the reorientation time increases until it reaches a point where the behaviours for both squirmers is different.
For the neutral squirmer, as we keep increasing  $\mathrm{Pe}$  the reorientation time decreases, reaching a minimum for the highest $\mathrm{Pe}$.
Whereas for the puller, at $\mathrm{Pe}=14.5$ there is a sharp decrease and then, as we keep increasing  $\mathrm{Pe}$, a slight recovery. Anyhow it is hard to draw  solid conclusions in both cases. One reason could also be due to not enough statistics.

\noindent
 Figure \ref{fig:OACF_chan}d-g offers a much clearer interpretation. In these panels we show how the OACF changes as we vary the channel radius keeping in all cases the maximum $\mathrm{Pe}$ available, corresponding to the highest thrust force $F_p=50$. As expected for a passive colloid (fig. \ref{fig:OACF_chan}.d) the OACF is the same regardless of the channel radius. Moving now to the pusher (fig. \ref{fig:OACF_chan}.e) we notice an increase of the reorientation time as we decrease the channel radius, consistent with the wall-rebound argument. For the puller (fig. \ref{fig:OACF_chan}.g) we encounter the opposite behaviour, the reorientation time increases with increasing radius, this can be explained with the wall-facing argument plus the fact that when the puller is swimming against the wall it is in an unstable state, similar to when a pencil is left standing at its tip, so it will change its orientation, some times this reorientation will lead him back to the center of the channel, but the narrower the channel, the sooner it will encounter again the wall and reorient again. For the neutral squirmer (fig. \ref{fig:OACF_chan}.f) we are again in between pushers and pullers but since neutrals propel slight in their front side, as pullers, the behaviour observed is more similar to pullers than to pushers.

\subsection{Active polymer}\label{subsec:polymer}

\noindent
In this section we present our results on 
structural and dynamical features of the active polymer in an explicit solvent. 
In particular, we focus on the radius of gyration $R_g$, and 
on the  diffusion coefficient $D$, computed  via the long-time behaviour of the  mean square displacement of the polymer's center of mass.  We consider the active polymer first in bulk and then confined in a cylindrical channel, underlying the effect of the activity in comparison with the passive polymer behaviour in the same conditions.  When in bulk, we unravel the effect of hydrodynamics  comparing our results to the results obtained in ref. \cite{bianco} for Active Brownian polymers (without hydrodynamics).\\

\subsubsection{Radius of Gyration}

\begin{figure*}[t!]
    \centering
    \includegraphics[width=0.8\textwidth]{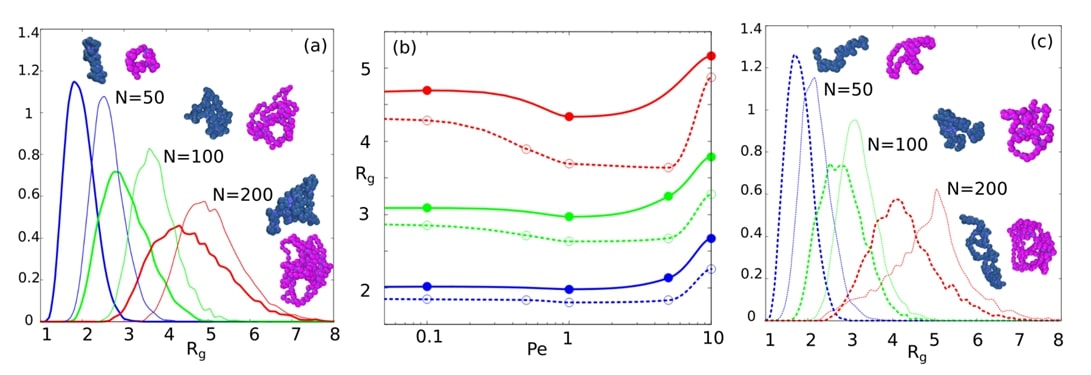}
    \caption{Probability distribution of the radius of gyration  for  passive ($\mathrm{Pe}=0$, thick line) and active ($\mathrm{Pe}=10$, thin line) polymers    of N=50 (blue), N=100 (green), N=200 (red) monomers in bulk (a) and in confinement (c). The  snapshots  illustrate typical conformations for the passive (blue) or active (magenta) polymers. (b) Average value of $R_g$ as function of $\mathrm{Pe}$ for different polymer sizes.  Continuous lines  are  for the polymer in bulk and dashed lines for the polymer under confinement. 
    }
    \label{fig:Rgpoly}
\end{figure*}

\noindent
Figure \ref{fig:Rgpoly} a and c shows the probability distribution function of the radius of gyration  for the polymer in bulk (panel a, continuous lines) and confined in a channel (panel c, dashed lines), comparing the passive (thick lines) to the active (thin lines) case. We  also sketch snapshots representing typical conformations  observed in each case, both for the passive case (in blue) and for the active one (in magenta). Panel b represents the average radious of gyration as a function of the P\'eclet number for polymer length of N=50 (blue), N=100 (green) and N=200 (red) in bulk (continous line) and in a channel (dashed line).

\noindent
When studying  a passive polymer in bulk (thick lines in fig. \ref{fig:Rgpoly}.a), our model recovers the expected  increase of the radius of gyration  with the polymer size. 
This  behaviour is observed also in the presence of active forces 
as shown by the thin lines in fig. \ref{fig:Rgpoly}.a.  In order to underline the relevance of hydrodynamics it would be interesting to compare the results obtained for the active polymer with those for the  Active Brownian Polymer reported in ref. \cite{bianco}. However, a direct comparison is not  possible, due to the different features of the chosen polymer's model.  In ref. \cite{bianco}  the authors used a bead-spring  self-avoiding polymer, whereas in our study we have used an ideal polymer.

\noindent
When studying the average of the radius of gyration as a function of the P\'eclet number (fig. \ref{fig:Rgpoly}-panel b) we detect a non-monotonic behaviour.
\noindent
For short polymers ($N=50$), $R_g$ remains constant at low activities (until $\mathrm{Pe}=5$): the same behaviour has been detected in ref. \cite{bianco} for active Brownian polymers, sign that hydrodynamics is not relevant for low activities.For relatively short polymers ($N=50$ and $100$)  the radius of gyrations is almost constant when activity is low activity,  and increases at high activity.
\textcolor{black}{This behaviour corresponds to what one would expect  if the polymer behaves like a flexible polymer \cite{Das2021}. The collapse is a consequence of the time scale separation of the thermal and active contributions. The subsequent  increase of the radius of gyration is due to the reduced influence of hydrodynamic interactions for larger values of Peclet number.}
\noindent
However, when activity increases, the presence of hydrodynamics affects the polymer conformation since $R_g$ increases. On the other side, without hydrodynamics $R_g$ decreases. For larger polymers ($N>50$) $R_g$ reaches a minimum value before increasing again. The same behaviour have been already reported in ref. \cite{Das2021} for active fully flexible Brownian self-avoiding polymer.
\noindent
Larger polymers ($N>200$) behave like a semi-flexible polymer \cite{Das2021}, characterised by an initial decrease of $R_g$ (more compact shape) for small values of the P\'eclet number, leading to an increase of $R_g$  with the activity (more open shape). This non-monotonic behaviour  resembles the behaviour observed for the end-to-end distance of active polymers in the presence of hydrodynamics\cite{winkler2020physics,eisenstecken2016conformational}.


\noindent
Even  when an active polymer is confined in a cylindrical channel (fig. \ref{fig:Rgpoly}.c), activity plays the same role on the probability distribution of the radius of gyration. The radius of gyration   increases with the number of monomers $N$ when hydrodynamics is taken into account. This is expected, as we increase the mass of a polymer.
Moreover, comparing the active (thin) to the passive (thick) polymer, the increase is more stretched in the active than in the passive case. Interestingly, the confinement does not seem to affect $R_g$ since same size active polymers ($ 50 \le N \le 200$) are characterised by the same radius of gyration when in bulk or in a channel. Probably, the reason for this is that we have chosen to study a channel whose diameter is relatively large, thus not differing too much from the bulk system. 








\subsubsection{Diffusion}

\begin{figure}[h!]
    \centering
    \includegraphics[width=\columnwidth]{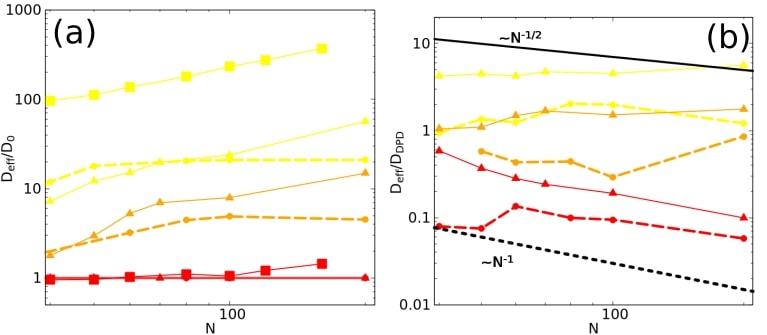}
    \caption{Diffusion coefficient of the polymer center of mass  $D_{eff}$ as a function of the polymer length for three different Pe values 0 (red), 0.1 (orange) and 1 (yellow). (a) Normalised  by the diffusion coefficient of the passive polymer ($D_0$); (b)   normalised by the  mean field diffusion of a DPD solvent particle Eq. (\ref{eq:diffusion_mean_field_DPD}). Square dots are results  from ref. \cite{bianco} with permission of the authors, triangles are results for the system in bulk and circles for system in confinement. 
    }
    \label{fig:diffpoly}
\end{figure}

\noindent
In order to understand the  dynamical features of an active polymer in bulk and in confinement, we compute the MSD of the active polymer's center of mass. 
\noindent
As in ref. \cite{bianco} for the system without hydrodynamics and like the squirmers studied in the previous section, MSD present at short time a ballistic regime ($MSD\propto t^2$), a diffusive dependence at long time ($MSD\propto t$) and for middle times there is a crossover characterize for a super-diffusive regime ($MSD\propto t^\nu$, with $1<\nu<2$). From the MSD  long time dependence, we estimate the diffusion coefficient.
Figure \ref{fig:diffpoly}  represents the diffusion coefficient $D_{eff}$  of the polymer's center of mass as a function of the polymer size, when varying the P\'eclet number. While in panel a we have normalised the diffusion coefficient by the diffusion coefficient of the passive polymer ($D_0$), in panel b we have normalised the diffusion coefficient by the mean field diffusion of a DPD solvent particle ($D_{sol}$ in eq.22). 

\noindent
In fig. \ref{fig:diffpoly}.a. we show the results for the effective diffusion normalized  by the diffusion coefficient of the passive case ($D_0$) as a function of the polymer size. We study values of activity ranging from the passive case (in red) to $\mathrm{Pe}=1$ (yellow case) and observe that  activity  increases the effective  polymer diffusion. Meanwhile, if we compare  the Brownian diffusion \cite{bianco} (yellow square) for the same activity $\mathrm{Pe}=1.0$ with our results (yellow triangles) we  detect the same dependence with $N$ but approximately 10 times smaller.
The effect of hydrodynamics is to slow down the polymers' motion, as expected. 
Finally, if we compare the results obtained for the bulk system with the ones for the channel we  observe how  confinement does not seem to affect the small polymers ($N<50$), but turns out to be relevant when the length of the polymer is increased. For longer polymers ($N>70$) the confinement affects  the polymer's  motion and the diffusion  decreases when the polymer is too long.

\noindent
On the other hand, in fig. \ref{fig:diffpoly}.b.  the effective diffusion has been normalized by the mean-field bead diffusion given by Eq. \ref{eq:diffusion_mean_field_DPD}. 
The idea to represent the data in this way was to be able to establish a power law dependence of the diffusion coefficient with the polymer size and compare it 
with the prediction expected for the diffusion by the Rouse and Zimm \cite{zimm1956dynamics} theory of Gaussian chains. Within this theory,  the chain center of mass diffusion $D_{Rouse} \propto N^{-1}$ and $D_{Zimm} \propto N^{-1/2}$. \textcolor{black}{As shown in  figure  \ref{fig:diffpoly}.b, when the polymer is passive (red line), the power law resembles that predicted by Zimm model, which take into account the hydrodynamic interactions between the beads of the polymer. While when the activity is relatively high (yellow line) the behavior is not similar to that expected in this model, then there are other effects due to the activity of the system.}

\noindent
\textcolor{black}{
In order to understand this behaviour, we compute the Schmidt number, defined as Sc=$\nu/D_{sol}$, being $\nu=1.25$ the kinematic viscosity of the DPD fluid and $D_{sol}$  the diffusion of the solvent particles. Estimating the kinematic viscosity via the mean field model of Groot and Warren’s  \cite{Groot}, and measuring  $D_{sol}$, we estimate Sc$=2.35$ for the solvent in all simulations. Whenever the Schmidt number is larger than one, the momentum diffusion dominates and hydrodynamics is relevant. However, since in our case the Schmidt number is around one, we conclude that the hydrodynamic coupling is not too strong. Therefore, we observe both scaling regimes, Rouse and Zimm.
}


\section{Discussion}\label{sec:discussion}

\noindent
In many paradigmatic examples of active matter such as biological microswimmers or synthetic active colloids, \textcolor{black}{these} are typically immersed in a solvent and the hydrodynamic interactions produced by the movement of the particles are relevant. Usually the introduction of these hydrodynamic interactions in active systems has been carried out through lattice models such as LB, that consider hydrodynamic effects but neglect thermal fluctuations, or MPCD that allow for the study of systems at low Reynolds number. In this work we develop a new \textcolor{black}{framework} for the introduction of hydrodynamic interactions in mesoscopic molecular dynamics simulations. To do so, we have used the well known DPD model that has been showed to be a simple and well behaved coarse-grain model (for an specific set of parameters) for the implementation of hydrodynamics interactions in passive systems. 

\noindent
One of the main advantages of this new implementation is the possibility of easily taking into account thermal fluctuations for swimmers of complex shapes.
Moreover, our implementation has been  developed as an extension of the  LAMMPSs open source package and will be sent to the LAMMPS developers (constantly maintaing the code), this makes our numerical approach readily available to be used for everyone. 

\noindent
In active systems there are a plethora of different mechanisms that produce the propulsion of the \textcolor{black}{microwimmers}, such as beating of flagella or chemical reactions. In our approach, we focus on the fact that in all these cases, the agents exert a force on the solvent in which they are immersed in order to achieve thrust. Depending on the type of propulsion mechanism  employed, the exerted force has its own distinct features but it always respects the  conservation laws of the different physical quantities. The model and its implementation is described in  details, based mainly on momentum conservation: this corresponds to the fact that the force experienced by the \textcolor{black}{microwimmer} in its propulsion must be compensated by the stresses induced in the solvent.

\noindent
To verify the validity of our model, we study two particular cases whose phenomenology has been well characterised by other numerical methods. The first of these cases are spherical squirmers, which represent the simplest model of a  \textcolor{black}{microwimmer} in which the hydrodynamics of the system is taken into account. 
The second example studied is an active polymer, which is nothing more than a first approximation to a slightly more complex structure: a chain of swimmers. In this case, our proposed method is applied in the same way for each of the monomers (swimmers) that form the polymer. 
As shown in the results section, the proposed method leads to a phenomenology, such as  flow fields,  dynamical and structural features, consistent with the results obtained for the same systems studied with different numerical models.

\noindent
Concerning the active colloid, we have been able to reproduce the solvent flow fields for the different types of swimmers (fig. \ref{fig:flow-fields_bulk}), observing a characteristic deformation of the solvent flow fields due to the inertial effects present in the fluid at moderate Reynolds numbers $\mathrm{Re}\sim 20$. We have been able to asses the impact of these inertial effects on the dynamics and hydrodynamics of the swimmer and to conclude that pushers are the most efficient swimmers (in the sense that they develop a larger propulsion velocity for the same propulsion force, fig. \ref{fig:prop_vel}), followed by neutrals and pullers when the Reynolds number is increased enough.  
For the ranges studied in the main part of our work, we showed that when swimming in bulk, diffusion is hardly affected by the choice of squirmer type (fig. \ref{fig:diff_col}a), and begins to saturate as we venture into higher Reynolds numbers. When the swimmer is confined inside a cylindrical channel, the flow fields changed dramatically in order for the fluid to adapt to the new boundary conditions (fig. \ref{fig:flow-fields_chan}). The confined geometry breaks this symmetry in the diffusion between the squirmer types (fig. \ref{fig:diff_col}b), as the behaviour of each swimmer near the channel wall is completely different: while pushers tend to rebound, aligning parallel to the wall and thus increasing their diffusion, pullers tend to get stuck, aligning perpendicularly with the wall and thus drastically decreasing their diffusion. Neutral squirmers lay in between both behaviours, but closer to pullers, as they slightly rely on the solvent ahead of them for achieving thrust. When varying the radius of the confining channel, diffusion of neutrals and pullers was hardly affected, while pushers enhanced their diffusion with decreasing channel radius (fig. \ref{fig:diff_col}c). Finally, we discussed the effect of the confinement in the reorientation time of the swimmers (fig. \ref{fig:OACF_chan}). We showed that pushers have slower reorientation dynamics the larger the P\'eclet number, but could not conclude anything solid for neutrals and pullers. Although when the P\'eclet is the highest and we increase the channel radius, the reorientation behaviour of pushers and pullers is clearly opposite: pushers increase their reorientation time while pullers decrease it. Again, neutrals lay in-between both behaviours although a little closer to pullers than to pushers.


\noindent
In the active polymer case, we have compared  the radius of gyration $R_g$ and diffusion to the system without hydrodynamics (Active Brownian Polymer). Concerning the radius of gyration $R_g$, the behavior of the polymer has been characterised as a function of both the polymer length and the Péclet number. 
Even though the radius of gyration monotonically increases  with the polymer length (fig. \ref{fig:Rgpoly}), the dependence with the activity is not so straightforward. For short polymers $R_g$ always increases with activity, whereas for long polymers it reaches a minimum value. This behaviour has been already detected in active fully flexible Brownian self-avoiding polymers. 
On the other hand,  confinement always decreases $R_g$ with respect to the system in  bulk. 
When studying  the dynamics of the polymer, we have compared our results with the analytical results for the Rouse and Zimm models, and concluded that  our model (for the set of parameters used) is compatible with the prediction of the Rouse model at low Peclet (when  hydrodynamics does not seem to  play a relevant role), and is compatible to the Zimm model at higher Peclet number, when the monomers of the polymer chain can interact  with each others due to  hydrodynamics.

\noindent
Having characterised the behaviour of individual swimmers in a solvent, we plan to use our numerical tool to  study more dense suspensions of active colloids or active polymers. This will allow us to study their collective behaviour, their aggregation (if present) and the interplay  played by  hydrodynamics and  activity, with the idea of comparing our numerical results on experiments on active synthetic colloids or active living swimmers (such as algae or bacteria), where hydrodynamics is relevant.



\section*{Acknowledgments}
C. Valeriani acknowledges fundings from MINECO PID2019-105343GB-I00 and EUR2021-122001 by  MCIN/AEI / 10.13039/501100011033 and FEDER, UE. I. Pagonabarraga acknowledges support from Ministerio de Ciencia, Innovaci\'on y Universidades MCIU/AEI/FEDER for financial support under grant agreement PGC2018-098373-B-100 AEI/FEDER-EU, from Generalitat de Catalunya under project 2017SGR-884, Swiss National Science Foundation Project No. 200021-175719 and the EU Horizon 2020 program through 766972-FET-OPEN NANOPHLOW.

\clearpage

\section{Appendix}\label{sec:Appendix}

\subsection{Detail on the hydrodynamic propulsion field}\label{subsec:hydro_details}
The radial and angular functions $f_r$ and $f_\theta$ can be expanded in terms of Legendre polynomials in a similar way than in the squirmer model, with the caveat that now we redistribute the forces on a spherical shell between $R_c$ and $R_H$ rather than just in the surface of the sphere,
\begin{align}
    f_r(r,\theta) &= P_{R_c,R_H}(r) \cdot \sum_n A_n P_n(\cos\theta)\\
    f_\theta(r,\theta) &= P_{R_c,R_H}(r)  \cdot \sum_n B_n V_n(\cos\theta)\\
    &\text{where}\,\, V_n(\cos\theta)=\frac{2\sin\theta}{n(n+1)}\frac{dP_n(\cos\theta)}{d\cos\theta}
\end{align}
where $P_{R_c,R_H}(r)=\Theta(r - R_c)\Theta(R_H - r)$ is a pulse function in the radial dimension which defines the spherical shell. Truncating the series for $n>2$ we arrive at,
\begin{align}
    f_r(r,\theta) &= \left[ A_0 + A_1\cos\theta+\frac{A_2}{2}(3\cos^2\theta-1)\right]\cdot P_{R_c,R_H}(r)\\
    f_\theta(r,\theta) &= \left(B_1\sin\theta+B_2\sin\theta\cos\theta\right)\cdot P_{R_c,R_H}(r)
\end{align}

\subsection{Propulsion velocity}\label{subsec:prop_vel}
Here we show the measured propulsion velocity of the colloid, $v_p=\langle\vec{v}\cdot\hat{e}\rangle_t$ with the corresponded P\'eclet and Reynolds numbers computed from eqs. \ref{ec:Peclet_col} and \ref{ec:Reynolds}, as a function of the propulsion force $F_p$ applied.
\begin{figure}[h!]
    \centering
    \includegraphics[width=\columnwidth]{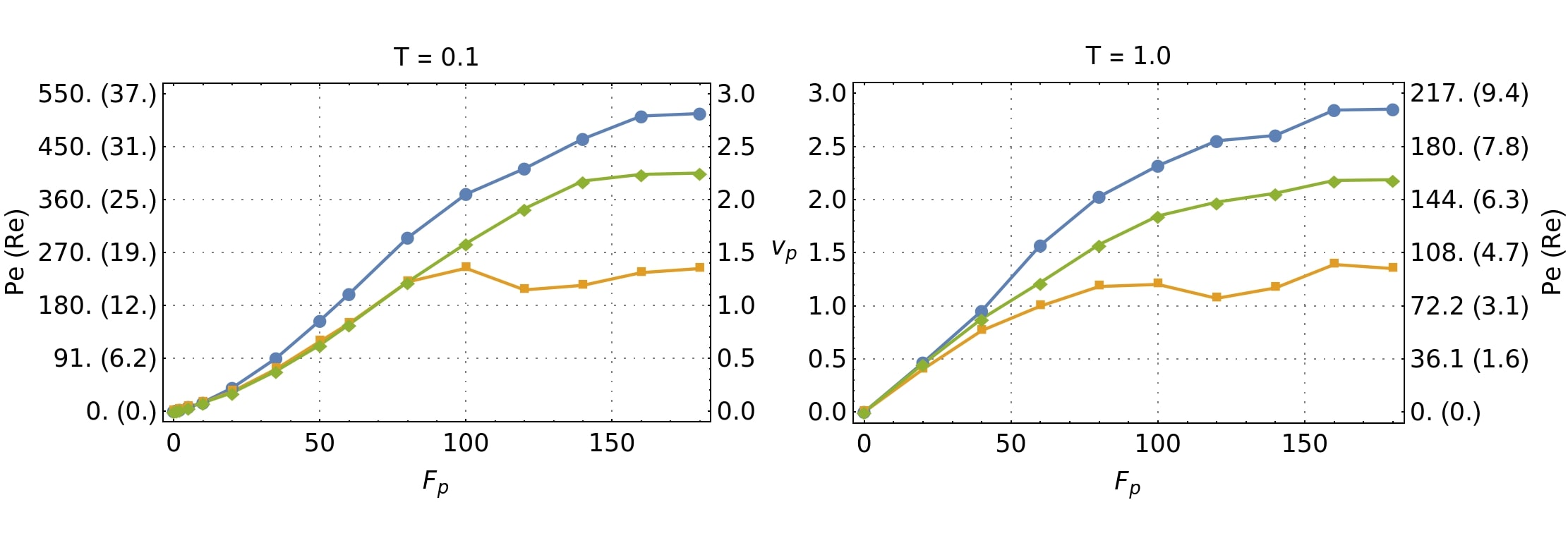}
    \caption{Propulsion velocity with its corresponding P\'eclet and Reynolds numbers as a function of the propulsion force of the colloid for the pusher (circles), neutral (diamonds) and puller (squares) squirmers, for $T=0.1$ (left) and $T=1.0$ (right). Note that for better readability, the right vertical axis of the left plot is shared with the left axis of the right plot and represents the propulsion velocity. The exterior axes represent the P\'eclet number and Reynolds number (between parenthesis) corresponding to each propulsion velocity.}
    \label{fig:prop_vel}
\end{figure}
As it can be seen the propulsion velocity is not linear with the propulsion force nor is the same for different kinds of squirmers. This maybe due to the fact that we are not at sufficiently low Reynolds number, since for low enough $\mathrm{Re}$ we notice that not only the relation is more linear but also that the curves collapse, giving almost the same values for the three types of squirmers. As a side result we can say that the pusher is the most efficient swimmer in this kind of environments, followed by the neutral and the puller squirmer.

\subsection{Low Reynolds number}\label{subsec:low_reynolds}
\begin{figure}[h!]
    \centering
    \includegraphics[width=\columnwidth]{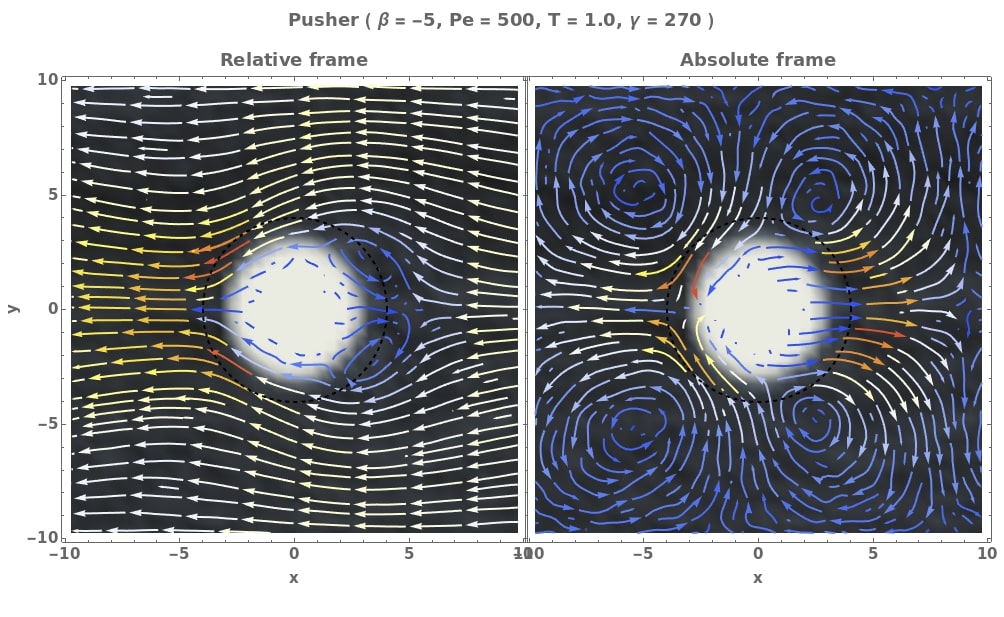}
    \caption{Pusher swimming to the right with $v_p\approx 0.4$, $\nu\approx 3.236 $, $\mathrm{Re}\approx 0.1236$, $\mathrm{Pe}\approx 659.58$. Here we can achieve this lower Reynolds number because we choose the friction coefficient of the DPD solvent as $\gamma^{ss}=270$.  \textbf{Left:} frame moving with the colloid. \textbf{Right:} lab frame. Compare with figs. \ref{fig:flow-fields_bulk}a and \ref{fig:flow-fields_bulk}d. }
    \label{fig:flow-field_lowRe}
\end{figure}

To verify that the deformation observed in the flow fields was indeed due to a high Reynolds number, we simulate a pusher in the same conditions as before but with a higher DPD friction coefficient ($\gamma^{ss} = 270$) between the solvent particles. The resulting flow field (fig. \ref{fig:flow-field_lowRe}) is less distorted (compare with figs. \ref{fig:flow-fields_bulk}a and \ref{fig:flow-fields_bulk}d), i.e. is more symmetric in the lab frame, with the vortices being more round, the saddle point at the front of the pusher is closer to the edge of the simulation box, where it should be ideally, and in the relative frame the swirls at the front of the pusher are more clearly visible. \textcolor{black}{The flow field for the absolute frame (fig. \ref{fig:flow-field_lowRe}, right) resembles more to the one obtained in ref. \cite{starkMPC,holmLBsquirmer}.}

\clearpage

\bibliographystyle{ieeetr}
\bibliography{biblio}



\clearpage

\end{document}